\documentclass[structabstract]{aa}  

\usepackage{graphicx}
\usepackage{txfonts}
\usepackage{natbib}
  \def\ProDiMo{{\sc ProDiMo\ }}

  \def\Td{T_{\hspace*{-0.2ex}\rm d}}
  \def\Tg{T_{\hspace*{-0.2ex}\rm g}}
  \def\cT2{c_T^2}
  
  \def\Rin{R_{\rm in}}
  \def\Rout{R_{\rm out}}

  \def\HH{{\rm\langle H\rangle}}
  \def\nH{n_{\HH}}

  \def\eg{e.\,g.\ }

  \def\amin{{a_{\rm min}}}
  \def\amax{{a_{\rm max}}}
  \def\apow{{a_{\rm pow}}}

\begin{document}

   \title{Radiation thermo-chemical models of protoplanetary disks}

   \subtitle{II. Line diagnostics}

   \author{I. Kamp
          \inst{1}
          \and
          I. Tilling\inst{2}
          \and
          P. Woitke\inst{2}
          \and
          W.-F. Thi\inst{3}
          \and
          M. Hogerheijde\inst{4}
          }

   \institute{Kapteyn Astronomical Institute, Postbus 800,
             9700 AV Groningen, The Netherlands
         \and
             UK Astronomy Technology Centre, Royal Observatory, Edinburgh,
              Blackford Hill, Edinburgh EH9 3HJ, UK
         \and
	     SUPA, Institute for Astronomy, Royal Observatory, Edinburgh,
              Blackford Hill, Edinburgh EH9 3HJ, UK
          \and
              Leiden Observatory, Leiden University, PO Box 9513, 2300 RA Leiden, The Netherlands
            }

   \date{Received ; accepted }

 
\abstract
   {}
    {In this paper, we explore the diagnostic power of the far-IR
     fine-structure lines of [O{\sc i}] 63.2$\,\mu$m, 145.5$\,\mu$m, [C{\sc
     ii}] 157.7$\,\mu$m, as well as the radio and sub-mm lines of CO
     J=1-0, 2-1 and 3-2 in application to disks around Herbig Ae stars. 
     We aim at understanding where the
     lines originate from, how the line formation process is affected
     by density, temperature and chemical abundance in the disk, and to
     what extent non-LTE effects are important.  The ultimate aim is to
     provide a robust way to determine the gas mass of protoplanetary
     disks from line observations.}
    {We use the recently developed disk code {{\sc ProDiMo}} to
     calculate the physico-chemical structure of protoplanetary disks
     and apply the Monte-Carlo line radiative transfer code {{\sc
     Ratran}} to predict observable line profiles and fluxes. We
     consider a series of Herbig Ae type disk models ranging from
     $10^{-6}$~M$_\odot$ to $2.2\!\times\!10^{-2}$~M$_\odot$ (between 
     0.5 and 700\,AU) to discuss the dependency of the line fluxes and
     ratios on disk mass for otherwise fixed disk parameters.
     This paper prepares for a more thorough multi-parameter analysis
     related to the Herschel open time key program {{\sc Gasps}}.}
    {We find the [C{\sc ii}]\,157.7\,$\mu$m line to originate in LTE from
     the surface layers of the disk, where $\Tg\neq\Td$. The 
     total emission is dominated by surface area and hence depends strongly 
     on disk outer radius. The [O{\sc i}] lines can be very bright
     ($>10^{-16}$~W/m$^2$) and form in slightly deeper and closer regions under
     non-LTE conditions. For low-mass models, the [O{\sc i}] lines come
     preferentially from the central regions of the disk, and the peak
     separation widens. The high-excitation [O{\sc i}]\,145.5\,$\mu$m
     line, which has a larger critical density, decreases more rapidly
     with disk mass than the 63.2\,$\mu$m line. Therefore, the [O{\sc i}]
     63.2\,$\mu$m/145.5\,$\mu$m ratio is a promising disk mass
     indicator, especially as it is independent of disk outer radius for
     $R_{\rm out}>200$~AU. CO is abundant only in deeper layers $A_V\!\ga\!0.05$.
     For too low disk masses ($M_{\rm disk}\!\la\!10^{-4}$~M$_\odot$) the
     dust starts to become transparent, and CO is almost completely
     photo-dissociated. For masses larger than that
     the lines are an excellent independent tracer of disk outer radius and can break the outer radius degeneracy in the [O{\sc i}]\,63.2\,$\mu$m/[C\,{\sc ii}]157.7\,$\mu$m line ratio.}
     {The far-IR fine-structure lines of [C{\sc ii}] and [O{\sc i}]
      observable with Herschel provide a promising tool to measure the
      disk gas mass, although they are mainly generated in the
      atomic surface layers.  In spatially unresolved observations, none of these lines carry much
      information about the inner, possibly hot regions $<\!30\,$AU.}



   \maketitle


\section{Introduction}

Observations of gas in protoplanetary disks are intrinsically difficult to interpret as they reflect the interplay between a complex chemical and thermal disk structure, statistical equilibrium and optical depth effects. This is particularly true if non-thermal excitation such as fluorescence or photodissociation dominate the statistical equilibrium.

The first studies of gas in protoplanetary disks concentrated on the rotational transitions of abundant molecules such as CO, HCN and HCO$^+$ \citep[e.g.][]{Beckwith1986, Koerner1993, Dutrey1997, vanZadelhoff2001, Thi2004}. Those lines originate in the outer regions of disks, $r>100$~AU, where densities are at most $n\sim10^7$~cm$^{-3}$. The interpretation of those lines was mainly based on tools and expertise developed for molecular clouds. Using the CO J=3-2 line, \citet{Dent2005} inferred for a sample of Herbig Ae and Vega-type stars a trend of disk outer radius with age; on average, the outer disk radius in the 7-20~Myr range is three times smaller than that in the $<7$~Myr range. Also, the disk radii inferred from the dust spectral energy distribution (SED) are generally smaller than those derived from the gas line \citep{Isella2007,Pietu2005}. \citet{Hughes2008b} suggest a soft outer edge as a solution to the discrepancy. Comparison of CO J=3-2 maps of four disks to different types of disk models strongly supports a soft edge in favor of a sharp cutoff. \citet{Pietu2007} use the CO and HCO$^+$ lines to probe the radial and vertical temperature profile of the disk. Simple power law disk models and LTE radiative transfer provides best matching results for radial temperature gradients around $r^{-0.5}$. The $^{12}$CO J=2-1 line, $^{13}$CO J=2-1 and $^{13}$CO J=1-0 lines  used in their analysis probe subsequently deeper layers and reveal a vertical temperature gradient ranging from 50~K in the higher layers to below the freeze-out temperature of CO in the midplane. This confirms earlier findings by \citet{Dartois2003}.

However, disk masses derived from the CO lines are in general lower than disk masses derived from dust observations \citep[e.g.][]{Zuckerman1995, Thi2001}. Possible explanations include CO ice formation in the cold midplane and photodissociation in the upper tenuous disk layers. Any single gas tracer alone can only provide gas masses of the species and volume from which it originates, the same way as dust masses derived from a single photometric measurement are only sensitive to grains of a particular size range, namely those grain sizes that dominate the emission at that photometric wavelength. Hence individual gas tracers are more valuable for probing the physical conditions of the volume where they arise than the total disk mass. A combination of suitable gas tracers can then allow us to characterize the gas properties in protoplanetary disks and study it during the planet formation process.

This paper aims at exploring the diagnostic power of the fine structure lines of [O\,{\sc i}] and [C\,{\sc ii}] in the framework of upcoming Herschel observations. Earlier modeling of these lines indicated that they should be detectable down to disk masses of $10^{-5}$~M$_\odot$ of gas, so also in the very gas-poor debris disks \citep{Kamp2005}. More recent work by \citet{Meijerink2008} presents fine structure lines from the inner 40~AU disk of an X-ray irradiated T Tauri disk; the models indicate that the [O\,{\sc i}] emission originates over a wide range of radii and depth and is sensitive to the X-ray luminosity. However, most of the C\,{\sc ii} and O\,{\sc i} line emission comes from larger radii where the gas temperature is dominated by UV heating processes. \citet{Jonkheid2007} find from thermo-chemical models of UV dominated Herbig Ae disks that the [O\,{\sc i}] lines are generally a factor 10 stronger that the [C\,{\sc ii}] line. We started to explore the origin of the fine structure lines in \citet{Woitke2009a} and find that the [C\,{\sc ii}]~157.7~$\mu$m line probes 
the upper flared surface layers of the outer disk while the [O\,{\sc i}]~63.2~$\mu$m line originates from the thermally decoupled surface layers inward of about 100~AU, above $A_V \approx 0.1$. The latter line is very sensitive to the gas temperature and might be used to distinguish between hot ($T_{\rm gas} \approx 1000$~K) and cold ($T_{\rm gas} = T_{\rm dust}$) disk atmospheres. Since the fine structure lines generally originate from a wider radial and vertical range than for example the $^{12}$CO rotational lines, they are potentially more suitable gas mass tracers.

We use in this paper the disk modeling code {{\sc ProDiMo}} presented in \citet{Woitke2009a} to study the gas line emission from disks around Herbig Ae stars. The main focus are the fine-structure lines of C\,{\sc ii} and O\,{\sc i} which will be observed for a large sample of disks during the {{\sc Herschel}} open time Key Program {{\sc Gasps}} (Gas evolution in protoplanetary systems: http://www.laeff.inta.es/projects/herschel). The disk parameters were chosen to resemble the disk around MWC480. According to previous work by \citet{Mannings1997b}, \citet{Thi2001} and \citet{Pietu2007}, the disk around this star extends from $\approx 0.5$~AU to 700~AU. We choose here a surface density profile $\Sigma \sim r^{-1.0}$. The central star is an A2e Herbig star with a mass of 2.2~M$_\odot$ and an effective temperature of $8500$~K.

Section~\ref{prodimo} gives a short summary of the disk modeling approach. The line radiative transfer method, re-gridding and the atomic input data are described in Sect.~\ref{LineRadTrans}. We then briefly discuss some basic properties of the Herbig Ae disk models (Sect.~\ref{resultsdisks}) before we present the fine-structure lines (Sect.~\ref{resultslines}) and conclude with a discussion of the diagnostic strength of fine structure line ratios and a comparison to previous ISO and submm observations (Sect.~\ref{discussion}).


\section{ProDiMo}
\label{prodimo}

\ProDiMo is a recently developed code for computing the hydrostatic structure of protoplanetary disks. This code  combines frequency-dependent 2D dust continuum radiative transfer, kinetic gas-phase and UV photo-chemistry, ice formation, and detailed non-LTE heating \& cooling with the consistent calculation of the hydrostatic disk structure. Details can be found in \citet{Woitke2009a}. We summarize in the following some other aspects that are particularly relevant to this study.

We use a {{\sc Phoenix}} stellar model \citet{Brott2005} with an effective temperature of 8500~K for the stellar irradiation and a highly diluted 20000~K black body for the IS radiation field that penetrates the disk from all sides. The dust opacities are computed using Mie theory and optical constants from \citet{Draine1984}. The chemical network contains 71 species (build from 9 elements) connected through 950 reactions (photo reactions, CR ionization, neutral-neutral, ion-molecule, as well as grain adsorption and desorption processes for CO, CO$_2$, H$_2$O, CH$_4$ and NH$_3$ ice). The gas temperature follows from an extensive heating and cooling balance that includes fine-structure line cooling as well as molecular and optical lines (fluorescence in the inner disk). The code does not yet include X-ray heating. This seems a minor issue for Herbig stars that generally have quite moderate X-ray luminosities, $L_X \ll 10^{30}$~erg~s$^{-1}$ \citep[e.g.][]{Stelzer2006}.

The treatment of photoionization and photodissociation in this work differs from that in the original code. 
The photoionization and photodissociation rates, $R^{\rm ph}$, are now computed at each grid point using the spectral photon energy density $\lambda u_\lambda$ calculated by the 2D continuum radiative transfer and the cross-sections $\sigma(\lambda)$ from the Leiden database \citep{vanDishoeck2008}. The photorate for  continuous absorption is
\begin{equation}
R^{\rm ph} = \frac{1}{h} \int \sigma(\lambda) \lambda u_\lambda d\lambda \,\,\, .
\end{equation}
When the photoprocess is initiated by line absorption, the rate becomes
\begin{equation}
R^{\rm ph} = \frac{\pi e^2}{mc^2}\lambda_j^2 f_j \eta_j (\lambda_j u_j/h)\,\,\, ,
\end{equation}
where $f_j$ is the oscillator strength for absorption from lower level $i$ to upper level $j$, $\eta_j$ is the efficiency of state $j$ with values lying between 0 and 1, and $\pi e^2/mc^2$ is 8.85 $\times$ 10$^{-21}$ with $\lambda$ in \AA.
The 2D UV radiative transfer uses now three bands defined between $91.2$, $111$, $145$, and  $205$~nm with central wavelengths of $100$, $127$, and $172$~nm and the radiation field at intermediate wavelengths is recovered through spline interpolation from the three bands.

We compute a series of Herbig Ae disk models with masses between $2.2 \,10^{-2}$ and $10^{-6}$~M$_\odot$. The parameters are summarized in Table~\ref{tab:Parameter}. The dust used in this model is typically larger than ISM dust and generates the $\sim \lambda^{-1}$ opacity law as derived from dust observations \citep[e.g.][]{Beckwith1991, Mannings1994, Rodmann2006}.

\begin{table}
\centering
\caption{Parameters of the Herbig Ae model series.}
\label{tab:Parameter}
\begin{tabular}{c|c|c}
\\[-4.5ex]
\hline
 Quantity & Symbol & Value\\
\hline 
\hline 
stellar mass                      & $M_\star$          & $2.2\,$M$_\odot$\\
effective temperature             & $T_{\rm eff}$      & $8500\,$K\\
stellar luminosity                & $L_\star$          &$32\,L_\odot$\\
\hline
disk mass                         & $M_{\rm disk}$     & $2.2\times 10^{-2}$, $10^{-2}, $ \\
                                           &                                 & $10^{-3}, 10^{-4}, 10^{-5},$\\
                                            &                               & $10^{-6}\,$M$_\odot$\\
inner disk radius                 & $\Rin$             & 0.5\,AU$^{\,(1)}$\\
outer disk radius                 & $\Rout$            & 700\,AU\\
radial column density power index & $\epsilon$         & 1.0\\
\hline
dust-to-gas mass ratio  & $\rho_d/\rho$      & 0.01\\
minimum dust particle radius      & $\amin$            & $0.05\,\mu$m\\
maximum dust particle radius      & $\amax$            & $200\,\mu$m\\
dust size distribution power index& $\apow$            & 3.5\\
dust material mass density        & $\rho_{\rm gr}$    & 2.5\,g\,cm$^{-3}$\\
\hline 
&&\\[-2.2ex]
strength of incident ISM UV       & $\chi^{\rm ISM}$   & 1\\
cosmic ray ionization rate of H$_2$  
                  & $\zeta_{\rm CR}$   
                  & $5\times 10^{-17}$~s$^{-1}$\\
abundance of PAHs relative to ISM & $f_{\rm PAH}$      & 0.1\\
$\alpha$ viscosity parameter      & $\alpha$           & 0.0\\
\hline
\end{tabular}\\[1mm]
\hspace*{0mm}\begin{minipage}{9cm}
\footnotesize
$(1)$: soft inner edge applied, see Sect.~3.1 of \citet{Woitke2009a}
\end{minipage}
\end{table}

\section{Line radiative transfer}
\label{LineRadTrans}

\subsection{Methods}

We use the two-dimensional Monte Carlo radiative transfer code {\sc Ratran} developed by \citet{Hogerheijde2000}. The code uses a two-step approach to solve the non-LTE line radiative transfer, {{\sc Amc}}, and {{\sc Sky}}. The first code solves the level population numbers for a given model atom/molecule within an arbitrary two-dimensional density and temperature distribution. The second one performs the ray tracing to derive the emission for a given line, distance and disk inclination. In the following, we discuss some aspects that are particularly relevant in applying these codes to complex chemical disk stratifications.

\subsection{Re-gridding}

  \def\Td{T_{\hspace*{-0.2ex}\rm dust}}
  \def\Tg{T_{\hspace*{-0.2ex}\rm gas}}
  \def\HH{{\rm\langle H\rangle}}
  \def\nH{n_{\HH}}

The 2D non-LTE line transfer code {\sc Ratran} requires a grid of
rectangular cells in cylindrical coordinates $r\!\in\!\{r_i,r_{i+1}\}$ and
$z\in\{z_j,z_{j+1}\}$ which is different from the grid of points used
in {\sc ProDiMo}. Therefore, we have to create a suitable grid of
cells for {\sc Ratran} and ``fill'' the cells in a physically sensible way, which will involve \
some kind of averaging for the physical quantities.

We choose to evaluate these mean values by integration and define the
following general function between the {\sc ProDiMo} grid points
\begin{equation}
  f(r,z) = f_0\,r^{\,p}\exp\big(-(z/H)^2\big) \ ,
\end{equation}
which can be applied to any physical quantity, \eg the species
particle density $n_{\rm sp}(r,z)$, or the product of hydrogen nuclei
density and gas temperature $(n_{\rm sp}\!\cdot\!\Tg)(r,z)$. Given any
point $(r,z)$ inside the {\sc ProDiMo} grid, we determine the indices
of the surrounding $2\times2$ {\sc ProDiMo} corner grid points, and
fix the free coefficients $f_0$, $p$ and $H$ to fit the values at
the corner points.

Next, we calculate the following integrals over the {\sc Ratran} cells
\begin{eqnarray}
  V_{ij}     &=& \pi (r^2_{i+1}-r^2_i) (z_{j+1}-z_j)\\
  N_{\rm sp,\,ij} &=& 2\pi
                 \int\limits_{r_i}^{r_{i+1}} r \int\limits_{z_j}^{z_{j+1}}
                 n_{\rm sp}(r,z) \,dz\,dr\\
  \langle N_{\rm sp}\Tg\rangle_{ij} &=& 2\pi
                 \int\limits_{r_i}^{r_{i+1}} r \int\limits_{z_j}^{z_{j+1}}
                 (n_{\rm sp}\!\cdot\!\Tg)(r,z) \,dz\,dr\\
  \langle N_\HH\Td\rangle_{ij} &=& 2\pi
                 \int\limits_{r_i}^{r_{i+1}} r \int\limits_{z_j}^{z_{j+1}}
                 (\nH\!\cdot\!\Td)(r,z) \,dz\,dr
\end{eqnarray}
to derive mean values for the species density, gas and dust temperature in our {\sc Ratran} cells
\begin{eqnarray}
  \overline{n_{\rm sp}}_{,\,ij} &=& N_{\rm sp,\,ij}\,/\,V_{ij}\\
  \overline{\Tg}_{,\,ij}        &=& \langle N_{\rm sp}\Tg\rangle_{ij}
                                    \,/\,N_{\rm sp,\,ij}\\
  \overline{\Td}_{,\,ij}        &=& \langle N_\HH\Td\rangle_{ij}
                                    \,/\,N_{\HH,\,ij}
\end{eqnarray}
Similar formulae apply to the collision partner densities.
We have choosen this procedure (i) to assure total and species mass
conservation, (ii) to conserve the total line emission in optically
thin LTE at long wavelengths (Rayleigh-Jeans approximation) and (iii)
to guaranty that the total thermal dust emission in the optically thin
case (Rayleigh-Jeans approximation) is conserved.

\begin{figure}
\centering
\includegraphics[width=8.5cm,height=6cm]{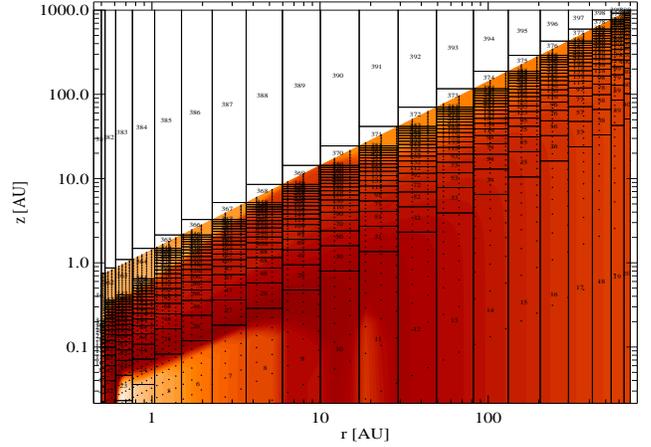}
\caption{Distribution of cells for the {{\sc Ratran}} radiative transfer code. The background contours show the O\,{\sc i} density distribution of the $10^{-2}$~M$_\odot$ model. The black boxes indicate the distribution of 400 cells across the model (original grid points are denoted by black dots).}
\label{fig:showgrid}
\end{figure}

Fig.~\ref{fig:showgrid} shows how the {\sc Ratran} grid compares to the original grid and underlying O\,{\sc i} density distribution ($10^{-2}$~M$_{\odot}$ model).

\subsection{Modified radiative transfer code {\sc Ratran}}\label{modified_ratran}

The original {\sc Ratran} code uses an accelerated Lambda iteration scheme to accelerate the convergence \citep{Hogerheijde2000}. For rather complex molecules with a large number of levels heavily interconnected through lines of various strengths, such as water, the performance is too slow to allow a full exploration of the disk parameter space. We describe in the following the implementation of two additional methods that significantly improve the {\sc Ratran} performance.

The line transfer simulations in {\sc Ratran} are carried out in two stages, called the "fixset" and the "random" stages. During the fixset stage the number, starting points and directions of the rays are fixed, and only the non-local feedback between level populations and spectral intensities are
solved iteratively. In order to accelerate the convergence during the
random-noise-free fixset stage, we have included the procedure of
\citet{Auer1984} for strictly convergent transformations, known as
the Ng-iteration. During the random phase, the number of rays is
successively increased in all cells until three different sets of rays
give approximately the same results.

The typical errors of the Monte-Carlo method depend on the chosen set of random numbers.
For standard (pseudo-random) number generators, the errors decrease with the number of photon packages as $N^{-1/2}$ \citep{Niederreiter1992}. Quasi-random numbers or low discrepancy sequences, to which the Sobol sequence belongs, try to sample more uniformly the distributed points
\citep{Sobol2007,Press2002}. The asymptotic errors of this quasi Monte-Carlo simulations 
decrease as $(\log(N))^d / N$, where $d$ is the dimension of the problem.  For {\sc Ratran}, the dimension is $d=5$, because the position in the cell, the direction and the Doppler velocity shift of the photon package are chosen randomly. The use of quasi-random sequences
was already advocated by \citet{Juvela1997}, since the gain in number of
photon packages is significant, but has not been widely adopted. One of the main critics of quasi-random number sequences is that there are only a limited number of sequences \citep{Sloan1993}. Hybrid
methods called randomized or scrambled quasi-random sequences combine
the advantages of both individual methods, infinite number of
sequences and low error, resulting in an asymptotic noise that
decreases in $N^{-s}(log(N))^{(d-1)/2}$, where $s$ is between 1 and
1.5 \citep{Hickernell2000}. Scramble quasi-random sequences show the
fastest noise decrease among the three types of sequences.

We replaced the pseudo-random sequence in  {\sc Ratran} by the quasi-random Sobol
sequence combined with a Owen-Faure-Tezuka type of scrambling
\citep{Owen2003}. A detailed description of the ACM
algorithm 659 implemented in {\sc Ratran} is given by \citet{Bratley1994}.
Difficult problems, like those involving water lines, require up to
$10^8$ photon packages with the standard random sequence but only
between 10$^4$ ($s=1.5$) and $\sim 2 \times 10^6$ ($s=1$) photon
packages with the scrambled quasi-random sequence, a gain in CPU time
between 50 ($s=1$) and $10^4$ ($s=1.5$). This speed gain opens the
possibility to study water line emissions from protoplanetary disks
with reasonable computing times \citep{Woitke2009b}.

We extended the original code also to include a larger number of collision partners. This proofs to be important as in the original code, the density of the first collision partner is used to compute the dust continuum emission. However, the density of one of the collision partners does not always equal the total hydrogen number density, thus causing inconsistencies either in the dust emission or the collision rates. Moreover, some lines originate over a wide range of physical and chemical conditions, so that the dominant collision partner changes between the various disk regions (e.g. from atomic hydrogen and electrons in the low density regions to molecular hydrogen and electrons deeper into the disk).

\subsection{Ray tracing}

We assume for our generic disk a distance of 131~pc and and inclination of 45 degrees. The pixel size (spatial resolution) is 0.05`` and the entire box has a size of  $13\times13$" to ensure that no emission is lost (disk diameter is 10"). The velocity resolution is set to 0.05~km/s and the total range is from -25 to 25~km/s for the C\,{\sc ii} line and -40 to 40~km/s for the O\,{\sc i} lines. The different velocity ranges reflect the differences in radial origin of the lines. For the oxygen fine structure lines, oversampling of the central pixels (out to 1.6") was used, i.e. an additional 2 rays generated per pixel.

The CO lines were computed within the same box of $13\times13$", but with a spatial resolution of 0.12" and a spectral
resolution of 0.2~km/s. No oversampling was used in this case.

\subsection{Atomic and molecular data}

Energy levels, statistical weights, Einstein A coefficients and collision cross sections are
taken from the Leiden {\sc Lambda} database \citep{Lambda2005}. Table~\ref{atomic_mol_data} 
provides an overview of the C\,{\sc II}, O\,{\sc I} and CO data used in this work. 

\begin{table}[htdp]
\caption{Atomic and molecular data taken from the {\sc Lambda} database  \citep{Lambda2005}.}
\begin{tabular}{lllll}
\hline
Species & \# Lev. & \# Lines & Collision & Reference\\
               &                &                & partner & \\
\hline
\hline
C$^+$ & 2 & 1 & H & \citet{Launay1977} \\
             &     &      & e$^-$ & \citet{Wilson2002} \\
             &     &      &  o-H$_2$ & \citet{Flower1977}\\
             &     &      &  p-H$_2$ & \citet{Flower1977}\\
O & 3 & 3 & H & \citet{Launay1977} \\
    &     &    & e$^{-}$ & \citet{Bell1998} \\
    &    &     & o-H$_2$ & \citet{Jaquet1992}\\
    &   &     & p-H$_2$ & \citet{Jaquet1992}\\
    &    &    & H$^+$& Chambaud (1980)$^{\rm (1)}$ \\
CO   & 26 & 25 & H & \citet{Chu1975} \\
        &       &       & H$_2$ & \citet{Schinke1985} \\
\hline
\end{tabular}\\[1mm]
\hspace*{0mm}\begin{minipage}{9cm}
\footnotesize
$(1)$: private communication
\end{minipage}
\label{atomic_mol_data}
\end{table}%

\begin{table}[ht]
\caption{Atomic and molecular data for the oxygen and carbon fine structure lines and the CO rotational lines.}
{\tiny
\begin{tabular}{lllllll}
\hline
 \multicolumn{3}{c}{Line identification} & $E_u$  & $E_l$  & $A_{ul}$  & $n_{\rm crit}$\\
 &  [$\mu$m] &  & [K] & [K]  & [s$^{-1}$] & [cm$^{-3}$]\\
\hline
\hline
O\,{\sc i} & 63.2   & $^3{\rm P}_1$ -- $^3{\rm P}_2$  &  227.7 & 0.0 & 8.865(-5) & 5(5)\\
O\,{\sc i} & 44.1   & $^3{\rm P}_0$ -- $^3{\rm P}_2$  & 326.6 & 0.0 & 1.275(-10) & 0.5 \\
O\,{\sc i} & 145.5  &  $^3{\rm P}_0$ -- $^3{\rm P}_1$ & 326.6 & 227.7 & 1.772(-5) & 6(4)\\
C\,{\sc ii} & 157.7  & $^2{\rm P}_\frac{3}{2}$ -- $^2{\rm P}_\frac{1}{2}$ & 91.2 & 0.0 & 2.300(-6) & 3(3)\\
\hline
 \multicolumn{3}{c}{Line identification} & $E_u$  & $E_l$  & $A_{ul}$  & $n_{\rm crit}$\\
 &  [GHz] &  & [K] & [K]  & [s$^{-1}$] & [cm$^{-3}$]\\
\hline
\hline
CO & 115. 3 & J = 1 -- 0 & 5.53 & 0.0 & 7.203(-8) & $^{(1)}$ 5.0(3) $T_{\rm g}^{-0.66}$ \\
CO & 230.5  & J = 2 -- 1 & 16.60 & 5.53 & 6.910(-7) & $^{(1)}$ 1.9(4) $T_{\rm g}^{-0.45}$\\
CO & 345.8  & J = 3 -- 2 & 33.19 & 16.60 & 2.497(-6) & $^{(1)}$ 4.6(4) $T_{\rm g}^{-0.35}$\\
\hline
\end{tabular}\\[1mm]
\hspace*{0mm}\begin{minipage}{9cm}
\footnotesize
$(1)$: CO critical densities from \citet{Kamp2001}\\
The notation 5(5) stands for $5\,10^5$.\\
Oxygen levels 2, 1, 0 correspond to $^3{\rm P}_2$, $^3{\rm P}_1$, and  $^3{\rm P}_0$, respectively.
\end{minipage}
}
\label{tab:oxygen}
\end{table}

\subsection{Dust opacities}

The dust opacities used in the radiative transfer code are consistent with the opacities used in the
computation of the disk model with {\sc ProDiMo}. The choice of opacities impacts the continuum around the line, i.e.\ the dust thermal emission. For oxygen, the low fine structure levels can be pumped by the thermal dust background. We see differences of up to 50\,\% in the continuum fluxes around the O\,{\sc i} fine structure lines and up to 20\,\% differences in the line emission itself when we choose either the grain size distribution from Table~\ref{tab:Parameter} using optical constants from \citet{Draine1984} or opacity tables from \citet{Ossenkopf1994}. The latter are based on an MRN size distribution for interstellar medium grains $f(a) \sim a^{-3.5}$ with sizes $5$~nm $< a <$ 0.25~$\mu$m. In this paper, the possibility of icy grain mantles or non-spherical shapes is not explored.

\section{The disk models}
\label{resultsdisks}

We compute a series of six disk models with different disk masses ($2.2\times 10^{-2}$, $10^{-2}$, $10^{-3}$, $10^{-4}$,  $10^{-5}$ and $10^{-6}$~M$_\odot$) and all other parameters remaining fixed. It is important to stress that this series of models is not intended to reflect an evolutionary sequence as we do not change the dust properties accordingly (e.g. dust grain sizes, dust-to-gas mass ratio, settling). The goal is to explore a range of physical conditions, study their impact on the disk chemistry and analyze how this impacts the cooling radiation that will be probed e.g. with the Herschel satellite.

\begin{figure*}
\centering
\includegraphics[width=18cm]{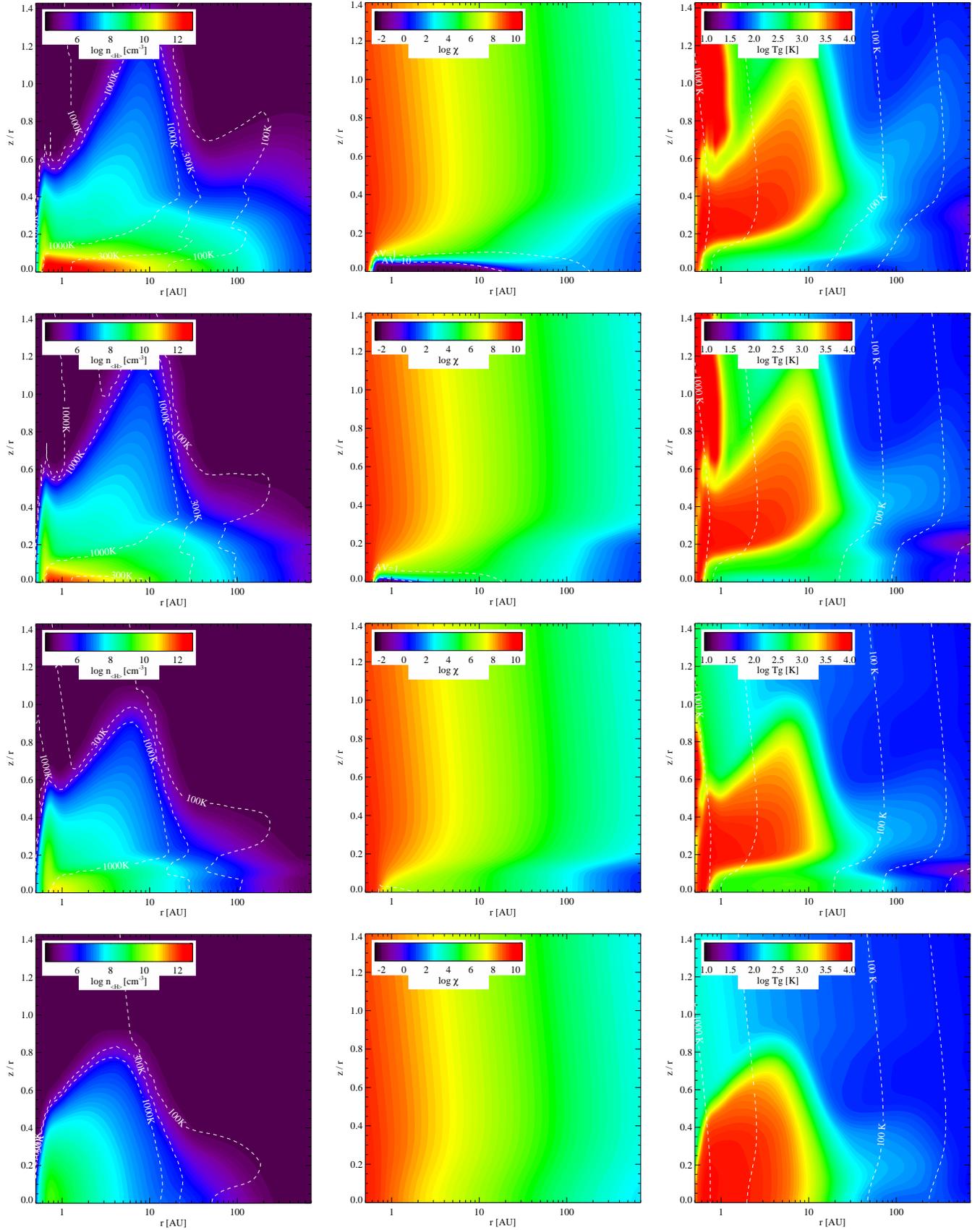}
\caption{From top to bottom row: disk models with $10^{-2}$, $10^{-3}$, $10^{-4}$ and $10^{-5}$~M$_\odot$. The first column shows the total hydrogen number density $\log n_\HH$ with white contours indicating gas temperatures of 100, 300, and 1000~K. The second column shows the UV radiation field strength $\chi$ ($91.2-205.0$ nm) from full 2D radiative transfer. The third column shows the gas temperature $\log T_{\rm g}$ with white contours indicating dust temperatures of 20, 40, 100, 300 and 1000~K. }
\label{fig:composite_nHChiTg}
\end{figure*}

\begin{figure*}
\centering
\includegraphics[width=18cm]{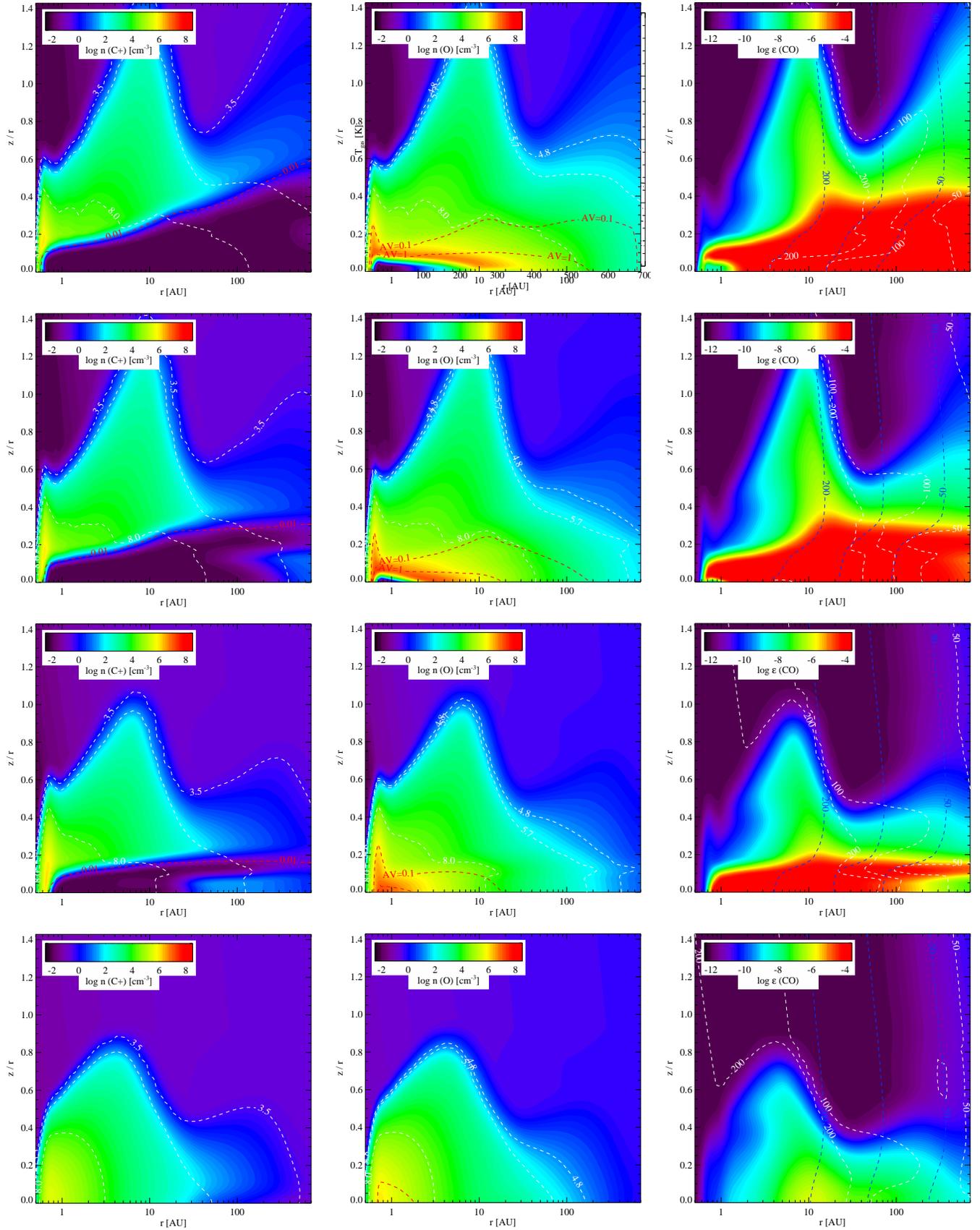}
\caption{From top to bottom row: disk models with $10^{-2}$, $10^{-3}$, $10^{-4}$ and $10^{-5}$~M$_\odot$. The first column shows the C$^+$ density with white contours indicating the critical density for the [C\,{\sc ii}] line and densities of $10^6$ and $10^8$~cm$^{-3}$ and a red contour line where the PDR parameter $\chi/n_\HH = 0.01$. The second column shows the O density with white contours indicating the critical densities for the two oxygen lines, $n_\HH = 6\,10^4$ and $5\,10^5$~cm$^{-3}$, and a density of $10^8$~cm$^{-3}$. Extinctions of $A_V=0.1$ and 1 are denoted by the red contour lines.  The third column shows the CO abundance with white contours indicating gas temperatures of 50, 100, and 200~K and the blue contours indicating dust temperatures of 50, 100, and 200~K). }
\label{fig:composite_nC+nOCO}
\end{figure*}

\subsection{Physical structure}
\label{physicalstructure}

\begin{figure*}
\centering
\includegraphics[width=18cm]{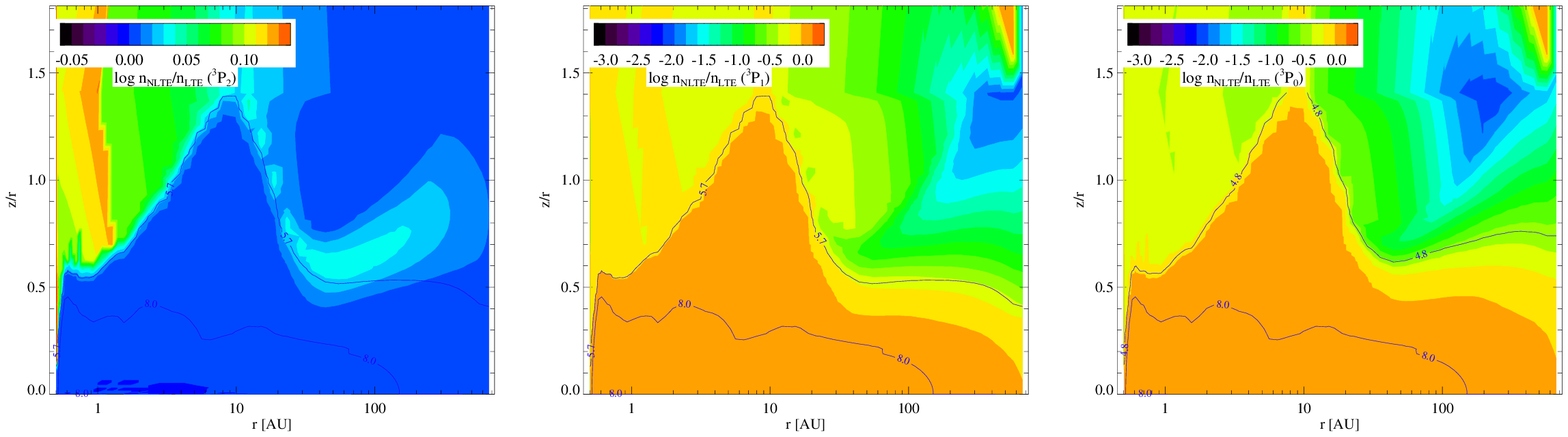}
\caption{Departure coefficient $\log n_{\rm NLTE}/n_{\rm LTE}$ for O\,{\sc i} in the $10^{-2}$~M$_\odot$ disk model. Blue contours denote the total hydrogen number densities with logarithmic values annotated in units of cm$^{-3}$.}
\label{fig:1e-2_levelpopsOI}
\end{figure*}

\begin{figure*}
\centering
\includegraphics[width=18cm]{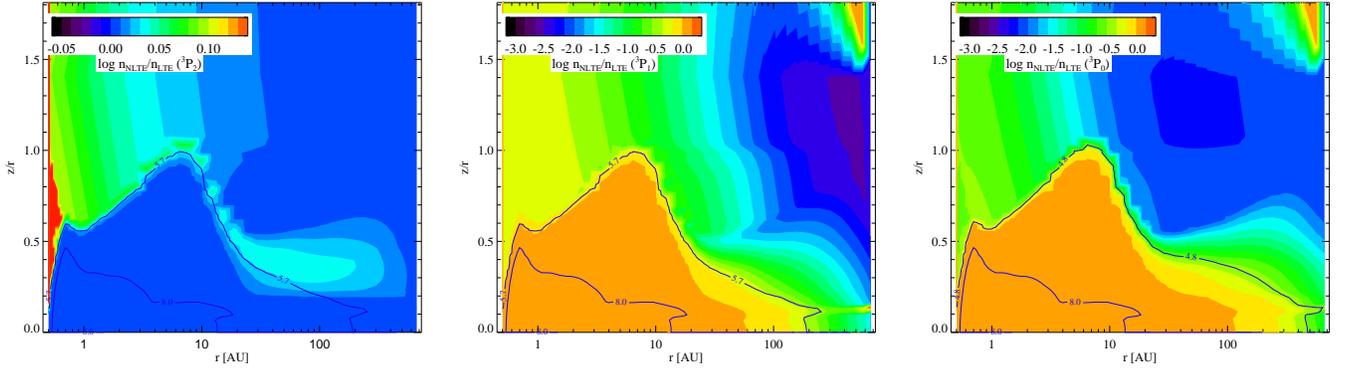}
\caption{Departure coefficient $\log n_{\rm NLTE}/n_{\rm LTE}$ for O\,{\sc i} in the $10^{-4}$~M$_\odot$ disk model. Blue contours denote the total hydrogen number densities with logarithmic values annotated in units of cm$^{-3}$.}
\label{fig:1e-4_levelpopsOI}
\end{figure*}

Fig.~\ref{fig:composite_nHChiTg} provides an overview of the computed physical quantities such as gas density, UV radiation field strength, optical extinction, gas temperature and dust temperature resulting from our series of Herbig disk models; each row presents a model with different mass and shows three quantities, total hydrogen number density, UV radiation field strength, and gas temperature. The white dashed lines in the second column of this figure indicate the $A_V=1$ and $A_V=10$ surfaces. The $2.2\times 10^{-2}$~M$_\odot$ model is optically thick in the vertical direction out to $\sim 400$~AU. Reducing the disk mass by a factor 200 leaves only a dense $\sim 10$~AU wide optically thick ring and the low mass disk models ($\leq 10^{-4}$~M$_\odot$) have very low vertical extinction, $A_V \lesssim 0.1$, throughout the entire disk. 

The high mass models show two vertically puffed up regions, one around 0.5~AU (the classical "inner rim"), the other one more radially extended between 5 and 10~AU. The very low mass models ($10^{-5}$~M$_\odot$ and below) do not show the typical flaring disk structure as observed in optically thick disk models, but they are still very extended in the vertical direction. The extreme gas temperatures in these low mass models are a result of direct heating by the photoelectric effect on small dust grains (PAHs) and pumping of Fe\,{\sc ii} by the stellar radiation field. In these optically thin disks, coupling between gas and dust grains is not efficient. 
 
\subsection{Chemical structure}

The particle densities of ionized carbon, oxygen and the abundance of CO are shown in Fig.~\ref{fig:composite_nC+nOCO}. The oxygen abundance is rather constant throughout the disks (C/O ratio $\sim 0.45$), being only lower by a factor of two where the CO and OH abundances are high; OH forms at high abundance inside 30~AU above an A$_V$ of a few, thus co-spatial with the warm ($> 200$~K) CO gas. Extremely low abundances occur only if significant amounts of water form close to the midplane inside 1-10 AU ($2.2\,10^{-2}$ -- $10^{-3}$ M$_\odot$ models). 

Carbon is fully ionized in the disk models with masses below $10^{-4}$~M$_\odot$. The C$^+$ abundance is complementary to the CO abundance as the neutral carbon layer between them is very thin. The C$^+$/C/CO transition is governed by PDR physics and can be well defined using the classical PDR parameter $\chi/n_\HH$. Above $\chi/n_\HH = 0.01$ (see Fig.~\ref{fig:composite_nC+nOCO}), carbon is fully ionized and its density structure resembles that of the total gas density (two puffed up inner rims). The column density of the ionized carbon layer is always smaller than a few times $10^{17}$~cm$^{-3}$. The mass of C$^+$ in the irradiated layers of the disk is roughly constant until the optically thin disk limit is reached in which case it is given by $M_{\rm gas} \epsilon({\rm carbon}) \left( m_{\rm C}/ \mu m_{\rm H} \right)$ (see Table~\ref{tab:masses}).

High abundances of CO can be found down to disk masses of $10^{-4}$~M$_\odot$. Below that, the entire disk becomes optically thin (radially and vertically), thus reducing the CO abundances even in the midplane to values below $10^{-6}$. In the models with masses above $10^{-3}$~M$_\odot$, densities in the innermost region are high enough to form a large water reservoir \citep{Woitke2009b}. At densities in excess of $\sim10^{11}$~cm$^{-3}$, the water formation consumes all oxygen, thus limiting the amount of CO that can form in the gas phase. In the upper disk layers, the CO abundance is very low due to the combined impact of UV irradiation from the inside (central star) and outside (diffuse interstellar radiation field). Fig.~\ref{fig:composite_nC+nOCO} also shows that the temperature of CO in the outer disk decouples from the dust temperature (white contours: gas, blue contours: dust), even though differences are generally within a factor two ($T_{\rm gas}\!>\!T_{\rm dust}$). This is relevant for the CO low rotational lines that are predominantly formed in those regions (see Sect.~\ref{sec:COsubmmlines}).

\begin{figure}
\centering
\includegraphics[width=9cm]{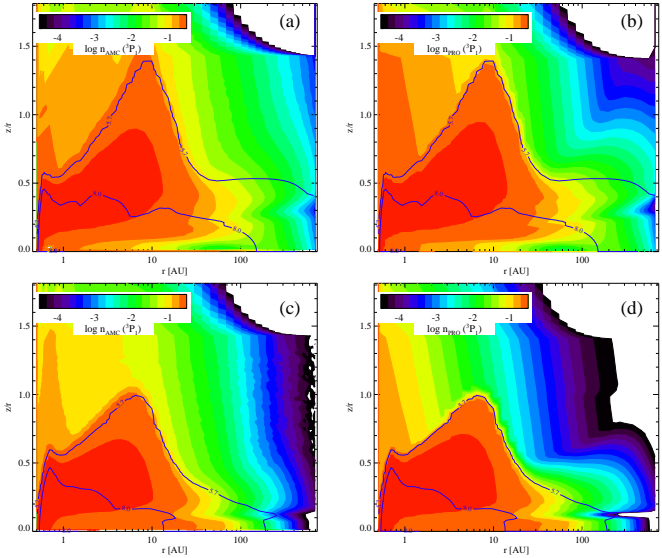}
\caption{O\,{\sc i} $^3$P$_1$ level population numbers from {{\sc Amc}}, $\log n_{\rm AMC}$, and escape probability (ES), $\log n_{\rm PRO}$, for the $10^{-2}$~M$_\odot$ disk model (a, b) and the $10^{-4}$~M$_\odot$ disk model (c, d). Blue contours denote the total hydrogen number densities with logarithmic values annotated in units of cm$^{-3}$.}
\label{fig:AMC_levelpopsOI}
\end{figure}

\begin{table}[ht]
\caption{Characteristics of selected species in the Herbig disk models. $\langle T_{\rm g} \rangle$ and $\langle T_{\rm d} \rangle$ are the mass averaged gas and dust temperature$^{(1)}$.}
\begin{tabular}{ll|lll}
\hline
M$_{\rm disk}$ [M$_\odot$] & Species & Mass [M$_\odot$] & $\langle T_{\rm g} \rangle$ [K] & $\langle T_{\rm d} \rangle$ [K\\
\hline
\hline
                   & O & 4.2(-5) & 61 & 34 \\
$2.2\,10^{-2}$ & C$^+$ & 1.3(-7) & 96 & 46\\
                   & CO & 6.0(-5) & 62 & 35 \\
                   \hline
                   & O & 2.1(-5) & 61 & 35 \\
$10^{-2}$ & C$^+$ & 1.1(-7) & 99 & 46 \\
                   & CO & 2.7(-5) & 61 & 35 \\
                   \hline
                   & O & 2.5(-6) & 64 & 34 \\
$10^{-3}$ & C$^+$ & 9.7(-8) & 79 & 34 \\
                   & CO & 1.8(-6) & 79 & 42 \\
                   \hline
                   & O & 3.2(-7) & 77 & 39 \\
$10^{-4}$ & C$^+$ & 4.9(-8) & 75 & 37 \\
                   & CO & 9.2(-8) & 103 & 51 \\
                   \hline
                   & O & 7.0(-8) & 92 & 50 \\
$10^{-5}$ & C$^+$ & 2.0(-8) & 91 & 48 \\
                   & CO & 3.6(-12) & 576 & 126 \\
                   \hline
                  & O & 5.1(-8) & 53 & 43 \\
$10^{-6}$ & C$^+$ & 1.7(-8) & 53 & 43 \\
                  & CO & 6.1(-15) & 89 & 54 \\
                   \hline
\end{tabular}\\[2mm]
\hspace*{0mm}\begin{minipage}{9cm}
\footnotesize
$(1)$: $\langle T \rangle = \frac{\int_{\rm V} T(r,z) n_{\rm sp}(r,z) dV}{\int_{\rm V} n_{\rm sp}(r,z) dV}$, 
species density $n_{\rm sp}$ and volume element $dV$.
\end{minipage}
\label{tab:masses}
\end{table}

\subsection{Self-similarity}

The six disk models show a large degree of self-similarity in their temperature and chemical structure. Lowering the disk mass removes mainly the thick midplane and the innermost dense regions. In that sense, this series of decreasing mass zooms in into disk layers further away from the star and at larger heights. The reason for this self-similarity lies partly in the two-direction escape probability used to compute the gas temperature and partly in the chemistry being independent of neighboring grid points (no diffusion or mixing). The solution in the disk is mostly described by the local $\chi/n_\HH$. 

\subsection{Strength of UV field}

To assess the impact of the UV field on the disk structure and line ratios, we computed two additional disk models with $10^{-2}$~M$_\odot$ disk masses, using effective temperatures of $9500$ and $10500$~K. This range reflects the typical temperatures encountered for Herbig Ae stars. To isolate the effect of UV irradiation, we keep the luminosity constant, so that a change in effective temperature changes only the fraction of UV versus optical irradiation. 

Increasing the stellar effective temperature to 10500~K leads to a vertically more extended disk structure, thus pushing the H/H$_2$ transition slightly outwards. Though the dust temperatures are unaffected (they depend rather on total luminosity), the mass averaged gas temperatures increase by up to a factor two for certain species such as CO and O. The change for C$^+$ being in the uppermost tenuous surface is more dramatic; its mass averaged temperatures increase from $\sim 100$~K ($T_{\rm eff}=8460$~K, Table~\ref{tab:masses}) to $330$~K ($T_{\rm eff}=10500$~K).

\subsection{Dust opacities}
\label{physicalstructure:dustopacities}

In a similar way, we varied the dust opacity in the $10^{-2}$~M$_\odot$ disk mass model, using first very small grains $a_{\rm min}=0.05$ to $a_{\rm max}=1$~micron and then only large grains $a_{\rm min}=1$ to $a_{\rm max}=200$~micron. They represent the two extremes of grain size distribution ranging from rather pristine ISM grains to conditions appropriate for more evolved dust in very old disks.

Dust opacities impact disk physics in two ways. First, increasing the average grain size decreases the opacity and thus the optical depth in the models. The dust temperature decreases mostly, except for the midplane regions inside 100~AU. However, the second --- more important --- effect is a decrease of effective grain surface area, thus decreasing the efficiency of gas-dust collisional coupling. As a consequence, the gas temperature decouples from the large grains even in the disk midplane, leading to higher gas temperatures everywhere and thus a vertically more extended disk. Given the fact that gas dust coupling is no longer efficient, the initial assumption that gas and dust are homogeneously mixed would have to be revisited. If most heating is provided by PAHs, this is not an issue as those will stay well mixed with the gas.

\section{Line emission}
\label{resultslines}

\begin{table*}
\caption{Results for the fine structure emission lines from various disk models and test
runs. In the model column, MC, LTE and ES denote the use of level populations from Monte Carlo line radiative transfer, LTE and escape probability, respectively. Line and continuum fluxes are shown for a distance of 131~pc and an inclination of 45 degrees. Numbers in parentheses indicate powers of ten, i.e. $1.93(3)$ denotes $1.93 \times 10^3$. The peak separation is defined as the distance between the two maxima in the double peaked line profile measured in km/s.}
{\tiny
\begin{tabular}{ll|lll|lll|lll}
\hline
& & \multicolumn{3}{|c|}{O\,{\sc i} 63\,$\mu$m} & \multicolumn{3}{|c|}{O\,{\sc i} 145\,$\mu$m} & \multicolumn{3}{|c}{C\,{\sc ii}~157.7\,$\mu$m} \\
M$_{\rm disk}$   & Model  & $F_{\rm line}$ & $F_{\rm cont}$ & Peak sep. & $F_{\rm line}$ & $F_{\rm cont}$ & Peak sep. & $F_{\rm line}$ & $F_{\rm cont}$ & Peak sep. \\
 ${\rm [}M_\odot{\rm ]}$ &  & [$10^{-18}$ W/m$^2$] & [Jy] & [km/s] &  [$10^{-18}$ W/m$^2$] & [Jy] & [km/s] & [$10^{-18}$ W/m$^2$] & [Jy] & [km/s] \\
\hline
\hline
$2.2\,10^{-2}$ & MC & 1.93(3) & 1.44(2) & 1.9 & 2.13(2) & 1.25(2) & 2.6 &     5.39(1) & 1.14(2) & 2.2 \\
$10^{-2}$ & MC & 1.29(3) & 1.03(2) & 2.4 & 1.18(2) & 6.57(1) & 3.0 &     4.69(1) & 5.91(1) & 2.2 \\
$10^{-3}$ & MC & 6.57(2) & 2.58(1) & 2.5 & 3.89(1) & 8.84 & 5.9 &     2.98(1) & 7.69 & 2.2 \\
$10^{-4}$ & MC & 2.65(2) & 4.56 & 5.9 & 8.06 & 9.09(-1) & 6.2 &     1.40(1) & 1.01 & 2.4 \\
$10^{-5}$ & MC & 4.19(1) & 9.89(-1) & 7.7 & 1.11 & 1.72(-1) & 6.0 &     4.53 & 2.71(-1) & 2.3 \\
$10^{-6}$ & MC & 1.99 & 1.50(-1) & 12.2 & 5(-2) & 3(-2) & 10.3 &    2.77 & 1.43(-1) & 2.2 \\ 
\hline
$10^{-2}$ & LTE & 2.25(3) & 1.05(2) & 2.2 & 1.30(2) & 6.63(1) & 2.7 &     5.22(1) & 5.96(1) & 2.2 \\
$10^{-2}$ & ES & 1.27(3) & 1.05(2) & 2.3 & 1.23(2) & 6.63(1) & 2.9 &     4.93(1) & 5.96(1) & 2.2 \\
\hline
$10^{-2}$ & no O\,{\sc i} $> 2000$\,K & 1.29(3) & 1.03(2) & 2.4 & 1.18(2) & 6.57(1) & 3.0 &     & & \\
$10^{-2}$ & $R_{\rm in,out}=0.5,500$\,AU & 1.23(3) & 1.06(2) & 2.6 & 1.14(2) & 6.16(1) & 3.2 &      2.26(1) & 5.41(1) & 2.7 \\
$10^{-2}$ & $R_{\rm in,out}=0.5,300$\,AU & 9.40(2) & 1.01(2) & 3.3 & 8.99(1) & 4.98(1) & 4.0 &      7.33 & 4.66(1) & 3.5 \\
$10^{-2}$ & $R_{\rm in,out}=0.5,100$\,AU & 5.74(2) & 8.27(1) & 5.6 & 4.18(1) & 3.19(1) & 6.0 &     1.35 & 2.76(1) & 6.3 \\
$10^{-2}$ & $R_{\rm in,out}=0.5,30$\,AU & 2.40(2) & 5.03(1) & 11.2 & 1.26(1) & 1.47(1) & 11.4 &     & & \\
$10^{-2}$ & $R_{\rm in,out}=30,100$\,AU & 3.70(2) & 3.42(1) & 5.5 & 3.08(1) & 1.75(1) & 6.1 &     & & \\
$10^{-2}$ & $R_{\rm in,out}=30,700$\,AU & 1.11(3) & 6.13(1) & 2.5 & 1.08(2) & 5.29(1) & 3.0 &     4.64(1) & 4.82(1) & 2.2 \\
$10^{-2}$ & $R_{\rm in,out}=100,700$\,AU &  & &  &  &  &  &     4.59(1) & 3.31(1) & 2.2 \\
\hline
$10^{-2}$ & $T_{\rm eff}=9500$~K & 5.53(3) & 1.68(2) & 2.8 & 4.56(2) & 8.60(1) & 2.9 & 1.50(2) & 7.63(1) & 2.3 \\ 
$10^{-2}$ & $T_{\rm eff}=10500$~K & 9.03(3) & 1.78(2) & 3.1 & 6.75(2) & 8.85(1) & 2.8 & 3.17(2) & 7.84(1) & 2.3 \\
\hline
$10^{-2}$ & $a_{\rm min,max}=0.01,1~\mu$m & 6.48(2) & 1.42(2) & 2.3 & 3.24(1) & 3.53(1) & 2.5 & 6.30(1) & 2.85(1) & 2.2 \\
$10^{-2}$ & $a_{\rm min,max}=1,200~\mu$m & 3.22(3) & 1.44(2) & 2.7 &  3.17(2) & 7.21(1) & 3.2 & 2.39(1) & 6.42(1) & 2.2 \\
\hline
\end{tabular}
\label{tab:OIlineemission_models}
}
\end{table*}

In the following, we performed radiative transfer calculations on the grid of Herbig Ae disk models to understand the spatial and physical origin of the two {{\sc Gasps}} tracers [C\,{\sc ii}], [O\,{\sc i}], and the frequently observed sub-mm lines of CO. Besides a wealth of published CO observations of protoplanetary disks to compare to and test the physics and chemical networks of our models, the low rotational CO lines are part of ancillary projects to complement the Herschel {{\sc Gasps}} project with tracers of the outer cold gas component in disks.

\subsection{[C\,{\sc II}]\,158\,$\mu$m line}

The fine structure line of ionized carbon arises in the outer surface layer of the disk. For values of $\chi/n_\HH$ smaller than 0.01, carbon turns atomic/molecular (see Fig.~\ref{fig:composite_nC+nOCO}, left column). This limits the C$^+$ column density and hence the [C\,{\sc ii}]\,158~$\mu$m line emission. Except in the very inner disk, $r<1$~AU, the line never becomes optically thick. 

\subsubsection{Line formation regions}

The line can be easily excited ($E_u=91.2$~K) even in the disk surface at the outer edge; hence the total [C\,{\sc ii}] emission from the disk is dominated by the 100-700~AU range and probes the gas temperature in those regions ($T_{\rm gas}\!\neq\!T_{\rm dust}$). At these distances, the column density of C$^+$ decreases with disk mass leading to a potential correlation of the [C\,{\sc ii}]\,158~$\mu$m line emission with total disk gas mass. However, as shown in Table~\ref{tab:OIlineemission_models}, the total line emission depends sensitively on the outer disk radius. The total [C\,{\sc ii}] line emission is thus degenerate for disk mass and outer radius.

In addition, the surrounding remnant molecular cloud material will also emit in the [C\,{\sc ii}] line, the only difference being in general lower densities and temperatures than those encountered in our protoplanetary disk models. The mass averaged gas temperature of C$^+$ in the disk is $\sim 90$~K. Densities range from up to $10^5$~cm$^{-3}$ in the outer disk (700 AU) to several times $10^8$~cm$^{-3}$ in the regions inside 10~AU (close to the  $\chi/n_\HH$=0.01 layer).

\subsubsection{LTE versus escape probability versus Monte Carlo}

Due to the low critical density of this line, the emission forms largely under LTE conditions. Deviations from LTE are small, less than 10\%, and grow towards lower disk mass models ($10^{-4}$ - 20\%, $10^{-5}$~M$_\odot$ - 40\%). Escape probability and Monte Carlo line fluxes agree well within 5-10\%, the typical uncertainty that can be expected from the re-gridding. Also here, the larger discrepancies are found in the lower mass disk models.

\subsection{[O\,{\sc I}]\,63 and 145\,$\mu$m lines}

Since our disk models span a much wider range of temperatures and densities than those found in molecular clouds and shocks, we encounter in this paper also different regimes for the formation of these two fine structure lines. The following paragraphs explore this in more detail.

\subsubsection{Line formation regions}

The column densities at which the 63~$\mu$m line becomes optically thick can be approximated ($n_l\sim n_{\rm O}$, $n_u \sim 0$) as
\begin{equation}
N_{\rm thick} = \frac{g_l}{g_u} \frac{8 \pi \nu^3 \Delta \nu_D}{A_{ul} c^3}
\end{equation}
where $g_u$ and $g_l$ are the statistical weights of the upper and lower level, $A_{ul}$ the Einstein A coefficient for the respective line transition with the frequency $\nu$. As an approximation, the Doppler width $\Delta \nu_D$ is assumed to be 1~km/s (within {\sc ProDiMo}, the Doppler width is calculated from the actual sum of thermal and turbulent broadening). The same formula can be used for the 145~$\mu$m line, if we take a factor into account correcting for the true level populations.  

The surface of the inner 30~AU of the $10^{-2}$~M$_\odot$ disk model is very hot with gas temperatures of several thousand K and typical densities above $10^7$~cm$^{-3}$. Thus, the relative level populations with respect to the total oxygen number density $n_{\rm O}$, in LTE, follow from the ratios of the statistical weights, $n_2/n_{\rm O} = 0.56$, $n_1/n_{\rm O}=0.33$, $n_0/n_{\rm O}=0.11$ (see Table~\ref{tab:oxygen} for the notation --- $n_J$ with $g=2J+1$ being the statistical weight of the level).  Under these physical circumstances, both O\,{\sc i} lines become optically thick at column densities of $\sim 3\,10^{17}$~cm$^{-2}$. Such column densities are reached inside 30~AU, even in our lowest mass disk model. However, the contribution to the integrated line emission from this hot gas is negligible. Removing the contribution from hot gas by setting the O\,{\sc i} abundance to zero for gas temperatures in excess of 2000~K, results in line flux changes that are less than 1\,\%.

In the 30-100~AU range, gas temperatures are much lower (few hundred K) and only about 10\,\% (3\,\%) of the oxygen atoms reside in the upper level of the 63 (145)~$\mu$m line. For those regions, the 145~$\mu$m line becomes optically thin in the $10^{-3}$ and $10^{-4}$~M$_\odot$ disk models. Emission from this ring dominates the total line emission. We will get back to this point later. Models that include X-ray heating and ionization \citep{Nomura2007,Meijerink2008} indicate that the impact of those additional processes is mostly relevant inside 30~AU. Beyond that, UV processes dominate the gas energy balance and chemistry. Hence, we do not expect X-rays to significantly alter the [O\,{\sc i}] line emission for disks with masses $> 10^{-4}$~M$_\odot$. In lower mass disks, the [O\,{\sc i}] emission originates closer to the star and here X-rays could have an impact by increasing the gas temperatures.

\begin{figure*}
\centering
\hspace*{-0.5cm}
\includegraphics[width=18cm]{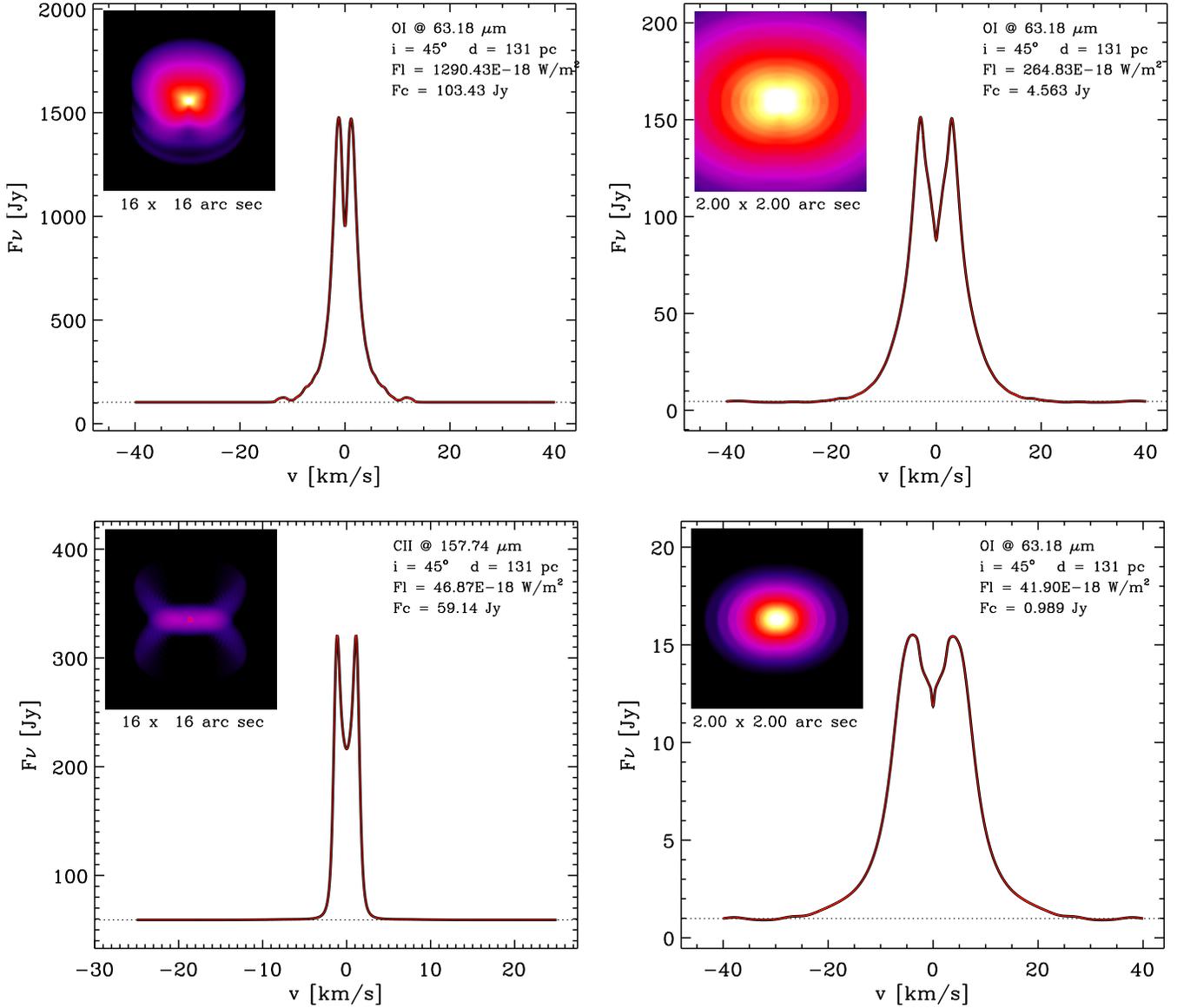}
\caption{Top row: [O\,{\sc i}]\,63\,$\mu$m line emission flux from the $10^{-2}$~M$_\odot$ and $10^{-4}$~M$_\odot$ model at a distance of 131~pc and an inclination of 45 degrees. The inserted image shows the continuum subtracted integrated logarithmic line intensity as a function of sky position. The angular size of 16" corresponds to 2100~AU at the distance of 131~pc. Bottom row: [C\,{\sc ii}]\,158\,$\mu$m line emission flux from the $10^{-2}$~M$_\odot$ model (note the different velocity range) and [O\,{\sc i}]\,63\,$\mu$m line emission flux from the $10^{-5}$~M$_\odot$ model. The color scale of the inserted image is the same in all panels ($I_{\rm max}=3\,10^{-11}$, $I_{\rm min}=3\,10^{-14}$ W/m$^2$/Hz/sr).}
\label{fig:OI63}
\end{figure*}

The outer disk, beyond 100~AU, is fairly cold ($T_{\rm g} < 200$~K). Here, level population numbers of the J=0 and J=1 level are extremely small (below 0.5\,\%). The peak separation $\delta {\rm v} = 1.9$~km/s of the [O\,{\sc i}]\,63~$\mu$m line in the $2.2\times 10^{-2}$~M$_\odot$ disk model (see Fig.~\ref{fig:OI63}) indicates that the emission comes mostly from inside 380~AU; the [O\,{\sc i}]\,145~$\mu$m originates slightly closer to the star, inside 200~AU ($\delta {\rm v} = 2.6$~km/s). The disk surface outside of 100~AU accounts for roughly 50\,\% of the O\,{\sc i} emission. For the lowest mass disk model, the physical conditions in this region such as gas/dust temperature and densities get close to molecular cloud values of $T_{\rm g}=50$~K,  $T_{\rm d}=40$~K, $n_\HH=10^5$~cm$^{-3}$.

In a series of radiative transfer calculations for the chemo-physical structure from the $10^{-2}$~M$_\odot$ disk model, we studied the fraction of the emission coming from regions with densities above $\tilde n_\HH$. Varying that density from the critical density of the 63~$\mu$m line, $\tilde n_\HH = n_{\rm crit} = 5\,10^5$~cm$^{-3}$, to a density of  $10^8$~cm$^{-3}$, shows that the regions below the critical density do not contribute very much to the total line emission. The emission gradually builds up with increasing density, showing that the line originates from regions up to an extinction of A$_V \sim 0.1$.

\subsubsection{LTE versus escape probability versus Monte Carlo}

LTE level population numbers systematically overestimate line fluxes by up to 70\%. In the regions where most of the O\,{\sc i} emission arises, the upper levels are significantly less populated than in LTE while the ground state is overpopulated with respect to LTE (Fig.~\ref{fig:1e-2_levelpopsOI} and \ref{fig:1e-4_levelpopsOI}). In the regions inside 30~AU, the H$_2$ abundance is low, thus the main collision partners are atomic hydrogen and electrons. Outside 30~AU, hydrogen is predominantly in molecular form.

The level population numbers computed with the simple escape probability assumption in \ProDiMo yield only slightly higher line fluxes, within 10\,\% for the O\,{\sc i} lines and $\sim 5$\,\% higher continuum fluxes. Calculating the line emission from Monte-Carlo radiative transfer requires a re-gridding of the  \ProDiMo model results. Fig.~\ref{fig:AMC_levelpopsOI} shows the level population numbers of the $^3$P$_1$ level of oxygen (upper level of the 63~$\mu$m line) for two models. Differences arise mainly in the very optically thin regions either at the disk surfaces or near the outer edge. In these areas, the maximum escape probability following our two directional escape probability approach is 0.5, while it should be rather 1.0 in very optically thin environments. Hence, in those areas the Monte Carlo approach gives a better result. 

Since continuum differences should largely be due to the re-gridding, the estimated error stemming from the interpolation onto a different grid is of the order of 5\,\%. Considering that, the intrinsic difference in line fluxes of both line radiative transfer methods is fairly small. The visible differences in the level population numbers are largest in the regions that do not contribute significantly to the total line emission (well above the critical density iso-contour in Fig.~\ref{fig:AMC_levelpopsOI}), i.e. the very tenuous surface and outer ($r \gg 100$~AU) regions of the disk.

\subsection{CO sub-mm lines}
\label{sec:COsubmmlines}

\begin{figure*}
  \centering
  \vspace*{-1mm}
  \begin{tabular}{ccc}
    \hspace*{-10mm}\includegraphics[height=5.0cm]{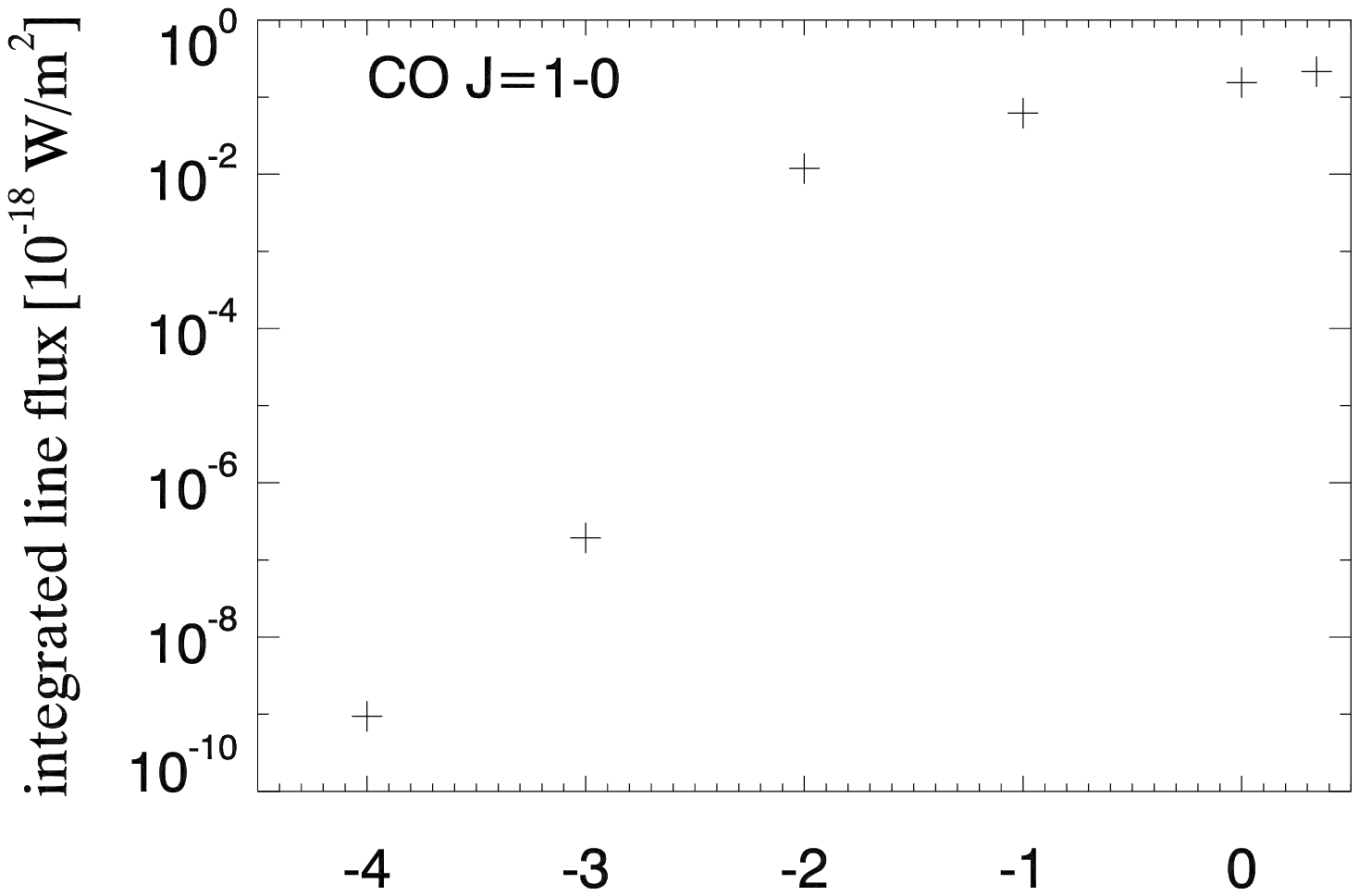} &
    \hspace*{-12mm}\includegraphics[height=5.0cm]{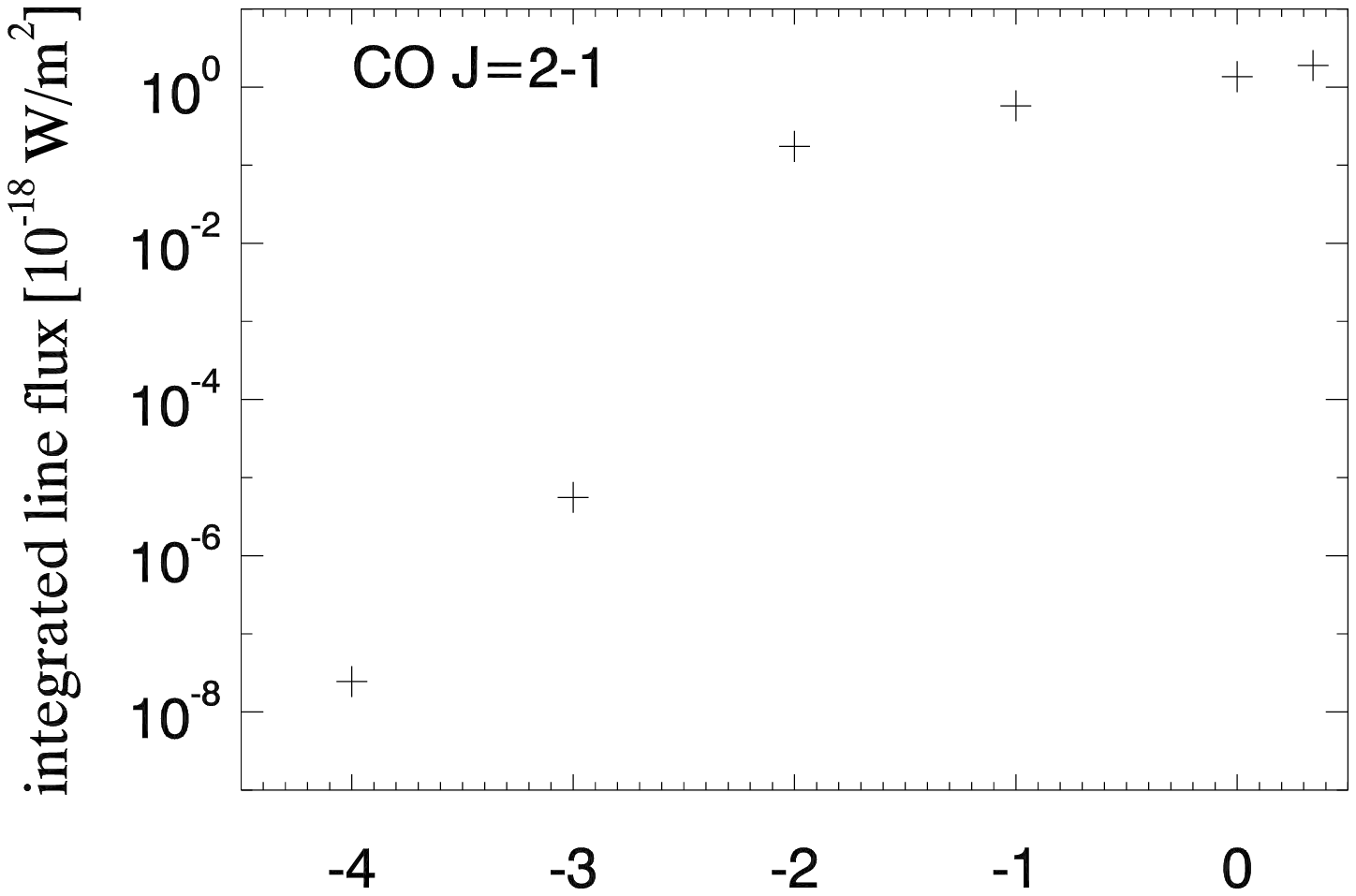} &
    \hspace*{-12mm}\includegraphics[height=5.0cm]{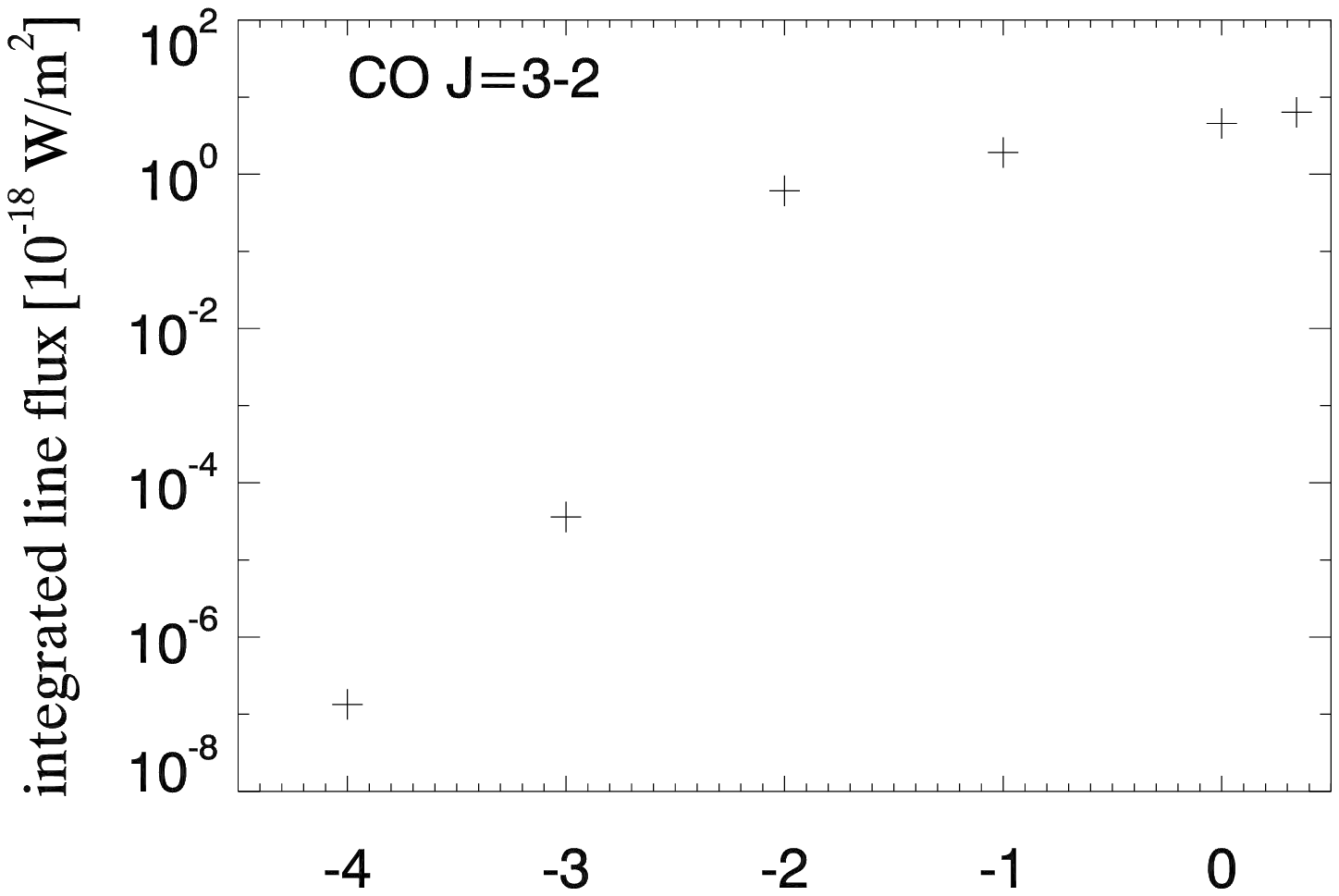}
    \\[-10mm]
    \hspace*{-10mm}\includegraphics[height=5.0cm]{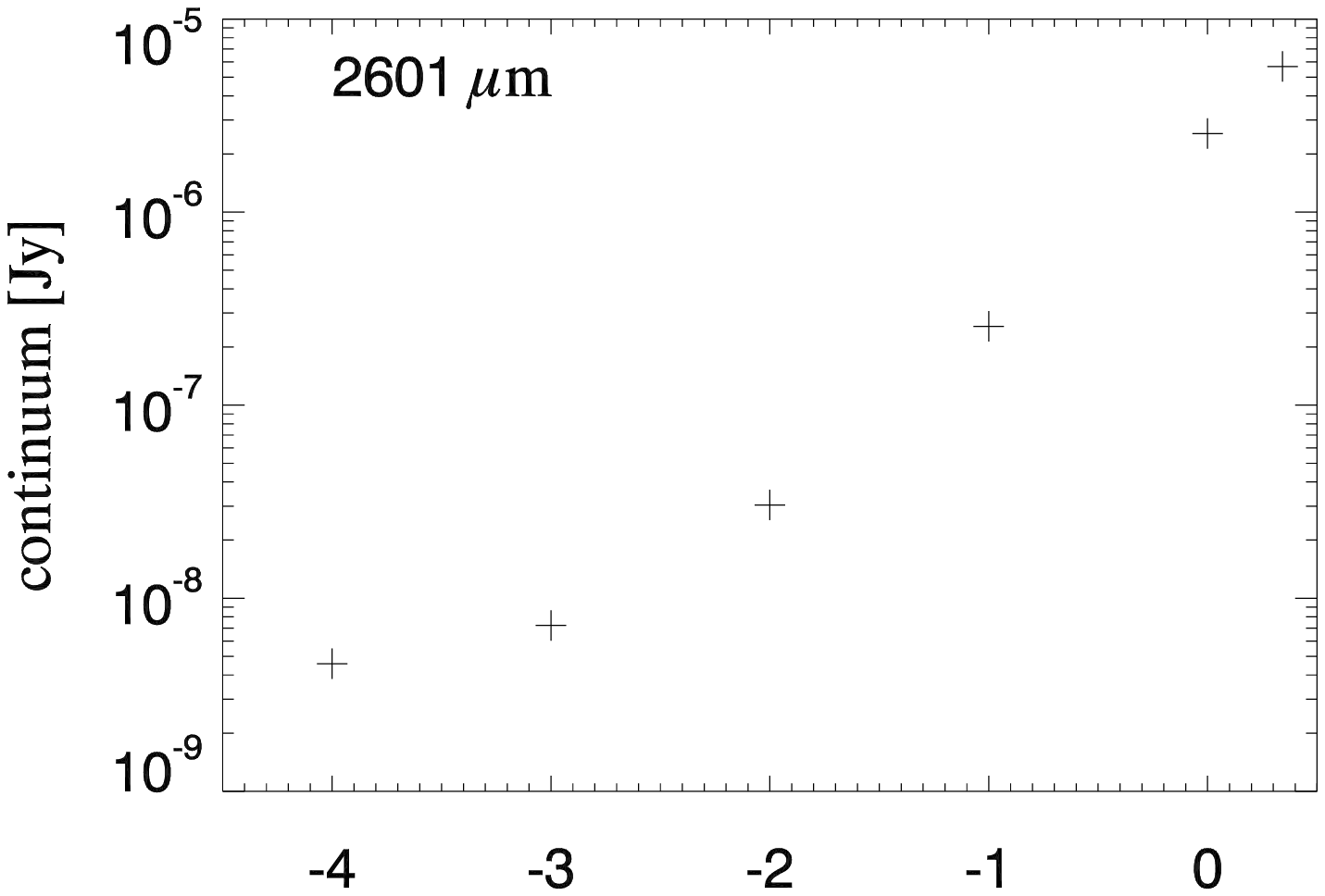} &
    \hspace*{-12mm}\includegraphics[height=5.0cm]{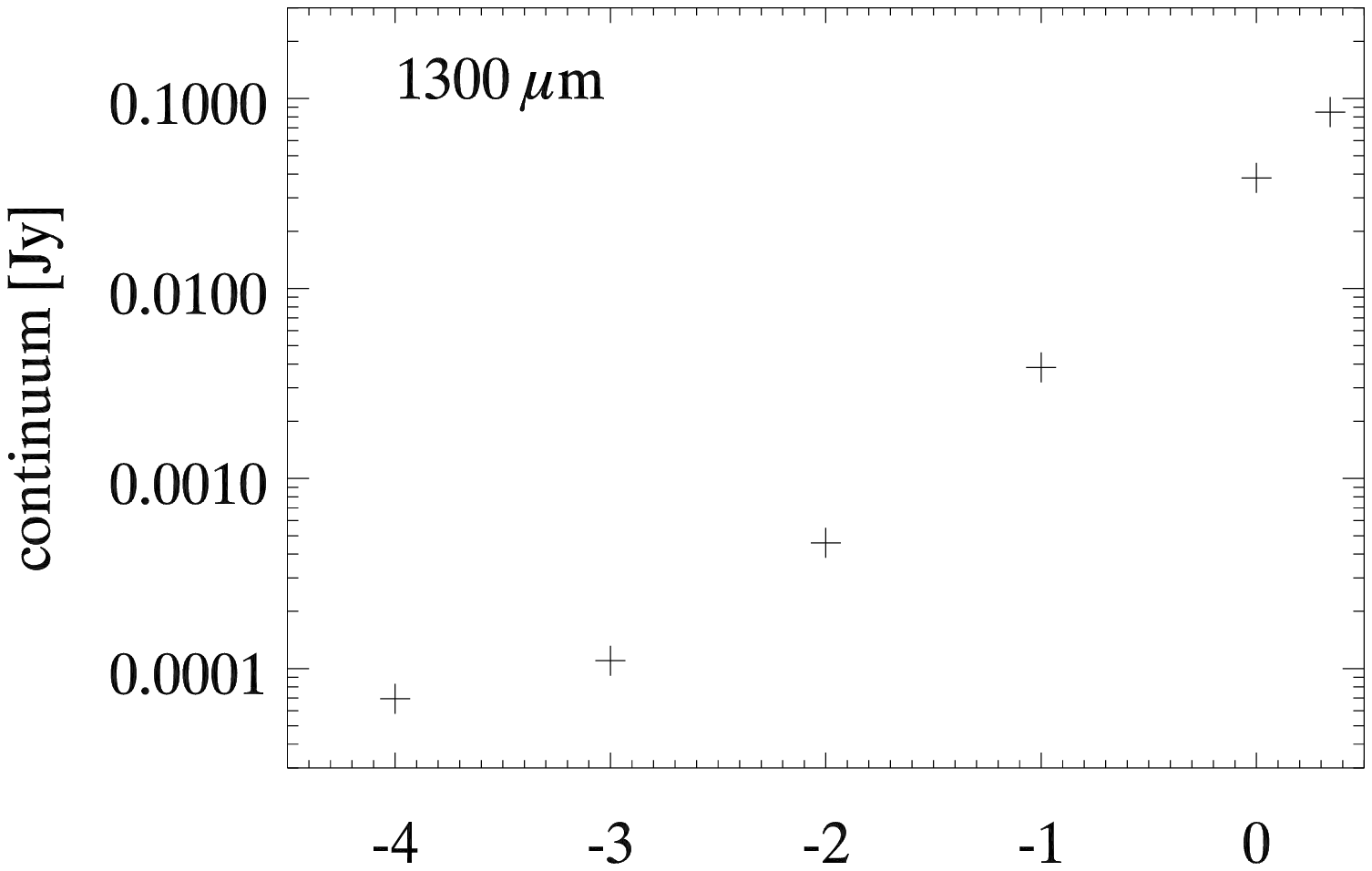} &
    \hspace*{-12mm}\includegraphics[height=5.0cm]{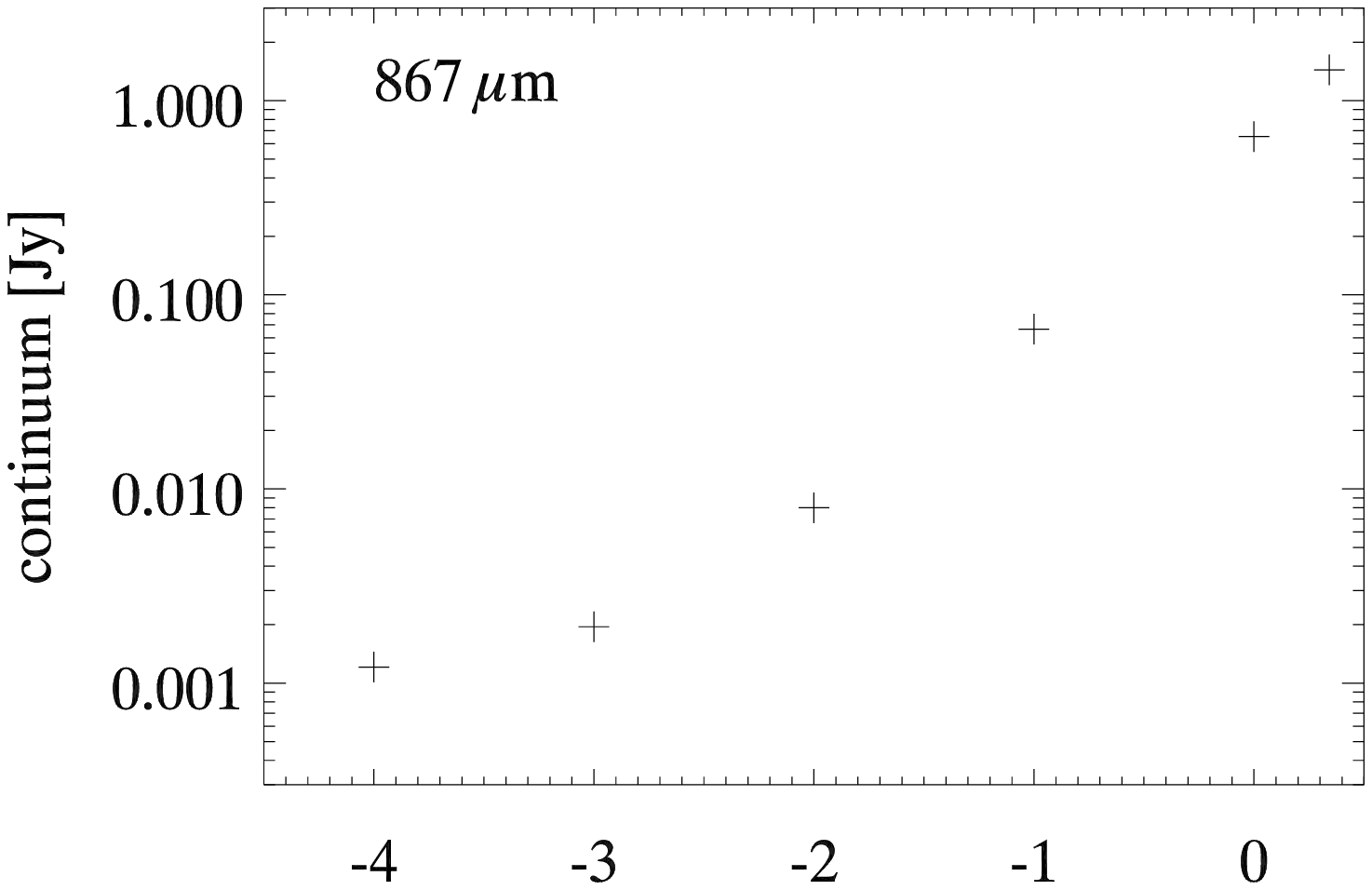}
    \\[-10mm]
    \hspace*{-10mm}\includegraphics[height=5.0cm]{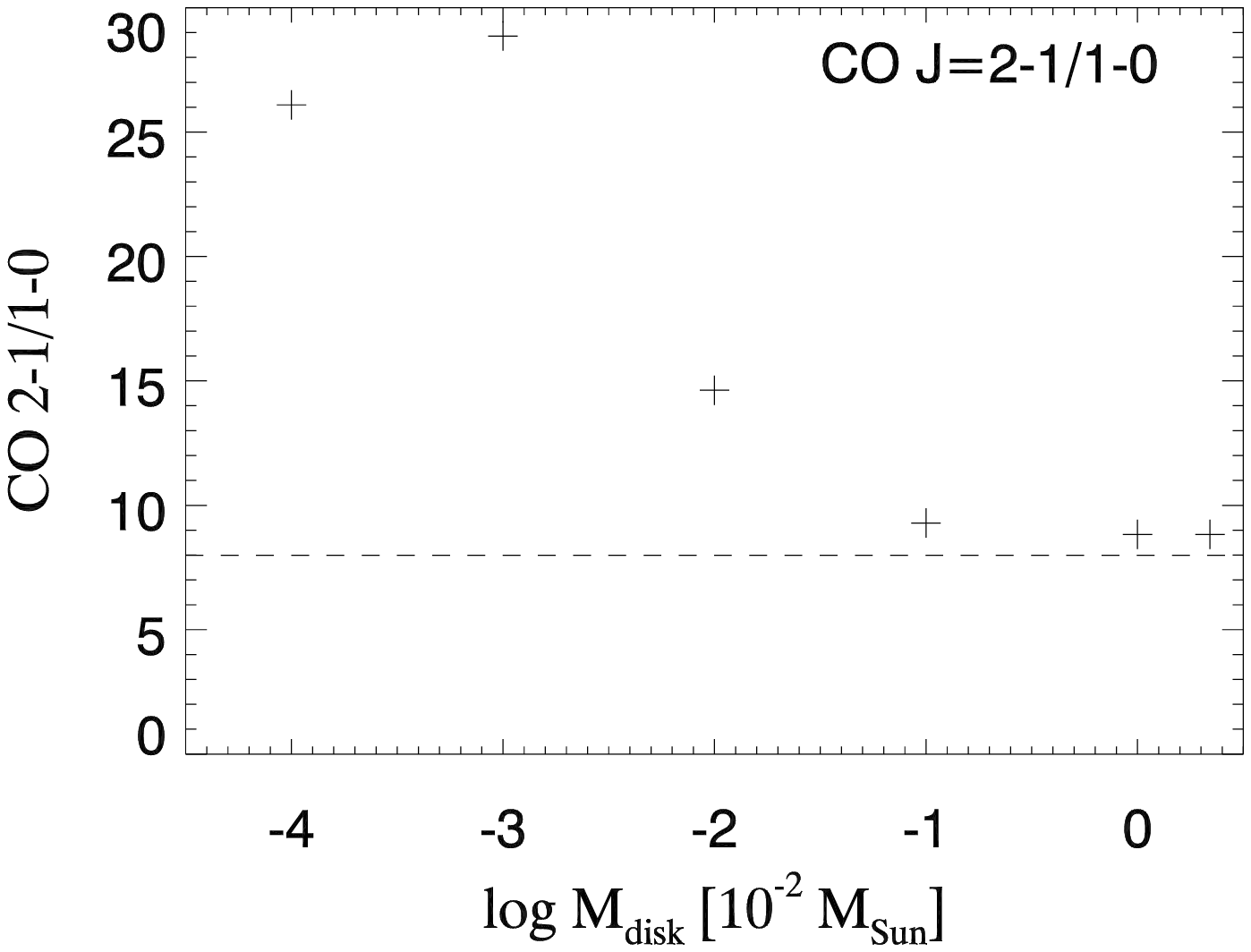} &
    \hspace*{-12mm}\includegraphics[height=5.0cm]{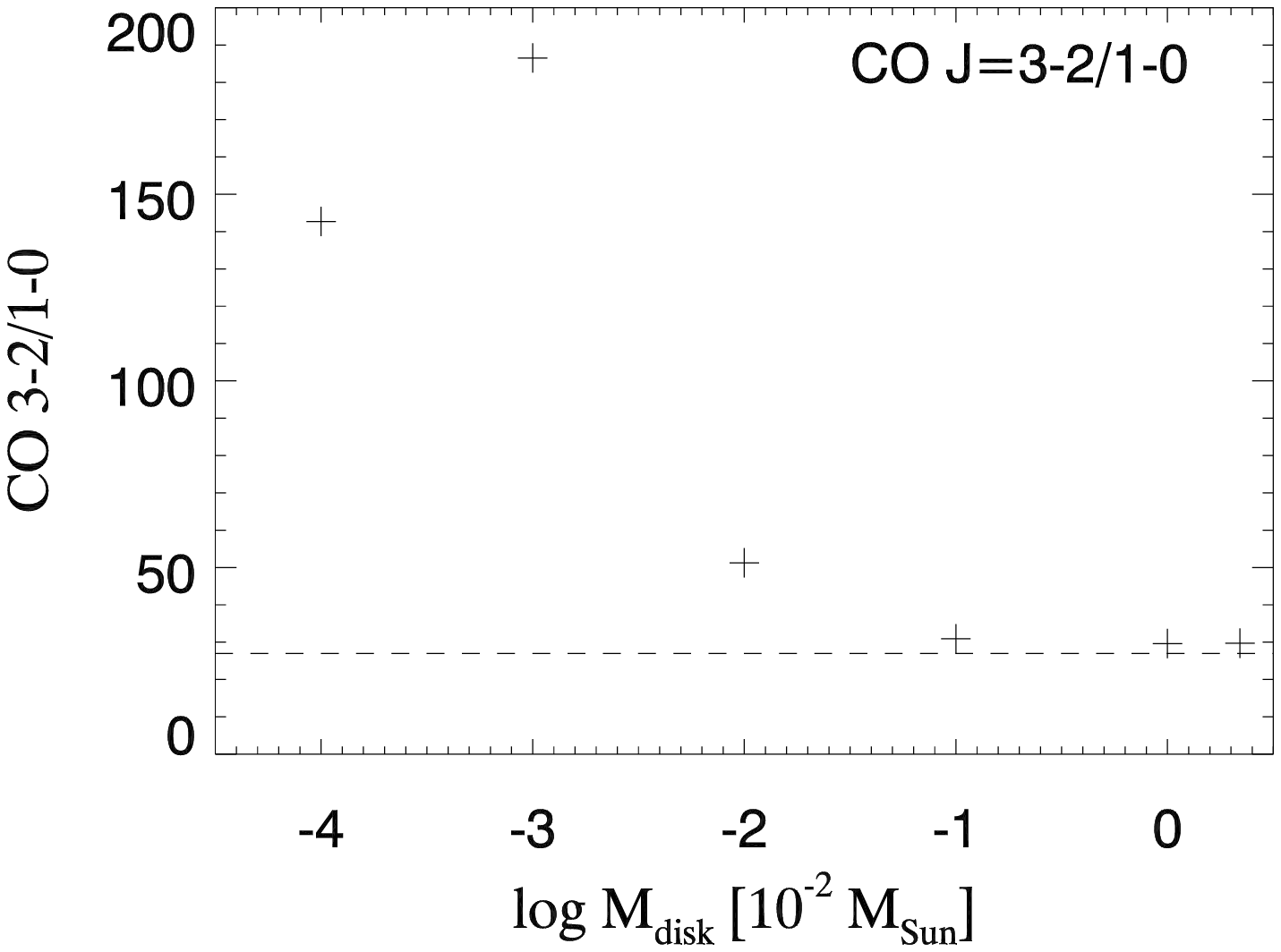} &
    \hspace*{-12mm}\includegraphics[height=5.0cm]{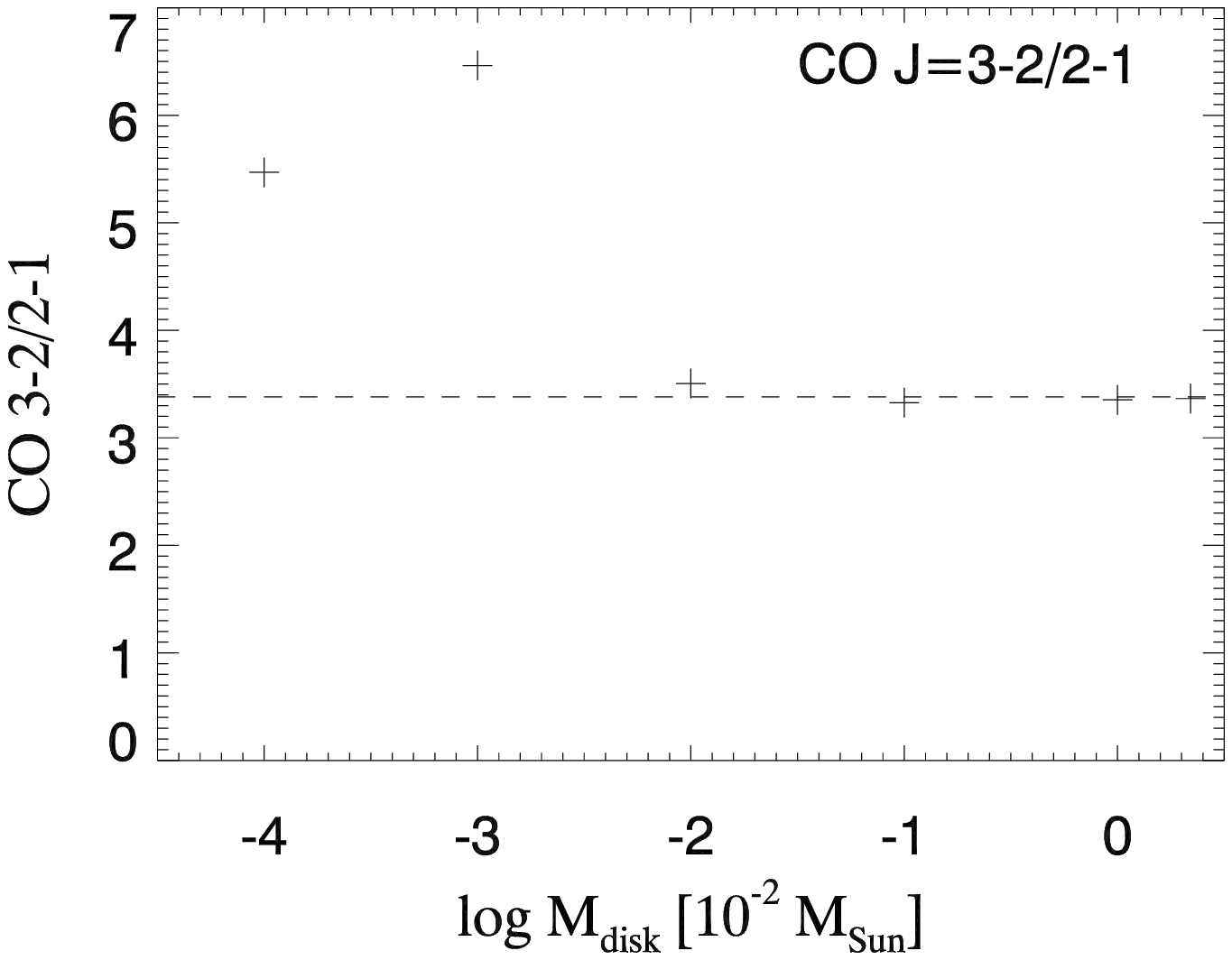}
  \end{tabular}\\*[-1mm]
  \caption{The top row shows integrated line fluxes as a function of
  disk mass, for (from left to right) the CO J=1-0, J=2-1 and J=3-2
  rotational lines. The second row shows the continuum fluxes at
  the corresponding line center wavelengths as a function of disk mass. 
  The third row shows the CO line ratios as a function of disk mass:(from left to
  right) J=2-1/1-0, J=3-2/1-0, J=3-2/2-1. Dashed lines indicate computed line ratios for optically thick lines in LTE.}
  \label{fig:co_fluxes}
  \vspace*{-2mm}
\end{figure*}

Line fluxes have been calculated for the first three molecular
transitions in CO, across the full mass range of disk models. The line
fluxes and continuum fluxes are plotted as a function of disk mass in
Fig.~\ref{fig:co_fluxes}. 

The J=1-0, J=2-1 and J=3-2 lines all show similar behaviour with disk
mass, with the line fluxes initially increasing sharply with mass
before levelling off for $M_{\rm disk} > 10^{-4}$~M$_{\odot}$. This
similarity results in largely uniform line ratios, with a spike at
$10^{-5}$~M$_{\odot}$. The continuum varies almost linearly with disk
(and hence dust) mass, although with slightly more emission from the lower mass models
than a strict linear relationship would give. This is due to the
optically thin low mass disks having more penetrating UV dust heating
in the midplane than the higher mass optically thick disks. 

The calulated line fluxes all exhibit a jump of four orders of
magnitude on moving from the $10^{-5}$ to the $10^{-4}$~M$_{\odot}$
model. This discontinuity is also seen in the line ratios, with all
three exhibiting a significantly higher ratio for the
$10^{-5}$~M$_{\odot}$ model than for the others. The sudden drop in emission below $10^{-4}$M$_{\odot}$
corresponds to a fall-off in $A_V$. There is suddenly almost no region
in the disk with $A_V > 0.1$, resulting in a maximum CO abundance of
less than $10^{-6}$ compared with $\sim 10^{-4}$ for the higher mass
models. The small remaining region with $A_V > 0.1$ is quite hot,
with gas temperatures around $500$~K, giving a spike in the line ratios at
$10^{-5}$~M$_{\odot}$. 

\subsubsection{Line formation regions}
\label{COlineform}

\begin{figure}
\centering
\includegraphics[width=9cm]{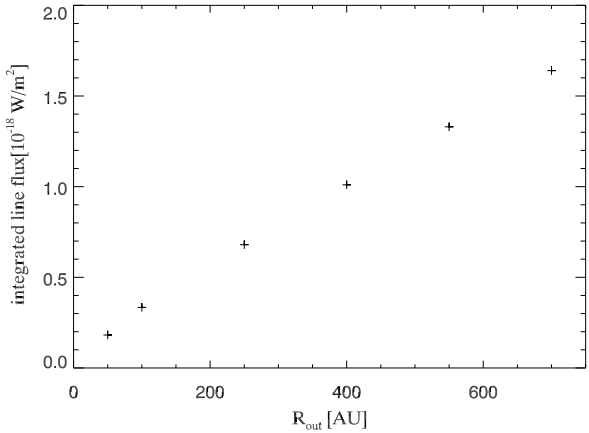}
\caption{Integrated CO J=3-2 line flux for the $10^{-3}$M$_{\odot}$
  model, plotted against the outer radius sampled by re-gridding.}
\label{fig:co_test}
\end{figure}

The three CO lines form at an intermediate height in the disk, between
the warm upper layer and the cold midplane \citep[see also][]{vanZadelhoff2001} and are
generally optically thick for disk masses down to $10^{-4}$~M$_\odot$. The submm line formation
region is radially very extended, with significant contributions to the total line
flux from the entire disk. The results of an experiment varying the
outer radius of the disk region sampled in the re-gridding procedure
are plotted in Fig.~\ref{fig:co_test}. The total J=3-2 line flux is seen to
vary linearly with $R_{\rm out}$ indicating that the line
originates from the full radial extent of the disk. The same behaviour
is seen in the J=1-0 and J=2-1 lines. The linear trend is caused by a combination of the radial 
gas temperature gradient in the CO emitting layers and surface area. The continuum shows a slightly different
behaviour, with a greater proportion of the integrated flux from outer
radii.

The line profiles for the three transitions are generally very similar for a given disk
model, and indeed across the computed mass range.  Narrow peak
separations ($\delta {\rm v}\!=\!1\!-\!2$~km/s) indicate that the emission is coming from the 
entire disk inside $\sim 700$~AU and will thus be dominated by the outer regions that
contain more surface area. 

\subsubsection{LTE versus escape probability versus Monte Carlo}

\begin{figure}
\centering
\includegraphics[width=9cm]{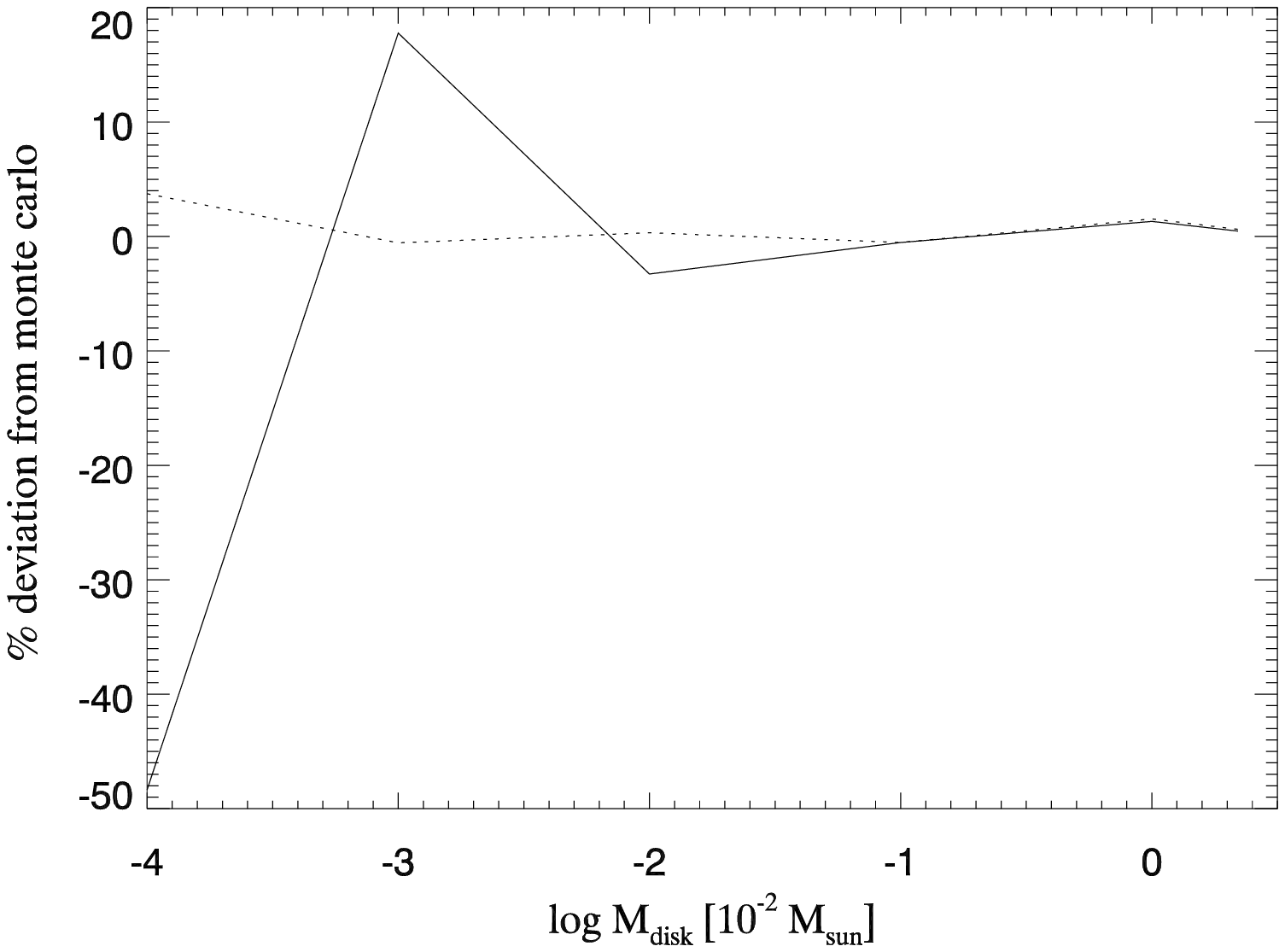}
\caption{Relative difference in percent between Monte Carlo CO line fluxes and CO line fluxes
calculated using the escape probabilty assumption (solid line) and between Monte Carlo CO line fluxes and LTE
CO line fluxes (dotted line); results are shown as a function of disk mass}
\label{fig:co_comparison}
\end{figure}

The continuum results obtained without re-gridding (ES) are in good agreement with those from the Monte
Carlo radiative transfer, within 3\% in all cases. This indicates that re-gridding does not present an issue when considering sub-mm fluxes. The escape probability line fluxes deviate slightly from the Monte Carlo and LTE fluxes for
the low disk masses, up to $\sim\!50$\% lower. This indicates again the limits of the two direction escape probability approach for the optically thin case of very tenuous disks. 

The escape probability fluxes are however in good agreement (within 3\%) for disk models with masses larger than$10^{-4}$M$_{\odot}$. The LTE line fluxes are also in good agreement (within $\sim1$\%) for the entire mass range (see Fig.~\ref{fig:co_comparison}), indicating that LTE is a valid approximation for these low CO lines. However, $T_{\rm gas}=T_{\rm dust}$ is not a valid approximation for this set of disk models, because the gas temperatures in the outer regions, where the CO lines arise, deviate by up to a factor two from the dust temperatures (see Table~\ref{tab:masses} for mass averaged gas and dust temperatures of CO). 

\subsubsection{CO line ratios}

To understand the values of the CO line ratios for high mass disks in Fig.~\ref{fig:co_fluxes}, we start with a number of assumptions: 1) The continuum is small $I_{\rm cont}\!\ll\!I_{\nu}$; 2) The CO line forms under optically thick LTE conditions with a universal line profile $\phi(\nu)$ such that $I_{\nu}\!=\!B_{\nu}(T_{\rm gas})\,\phi(\nu)$ where $B_{\nu}$ is the Planck function; 3) The temperature $T_{\rm gas}$ is constant throughout the line forming region. The line flux can then be expressed as
\begin{equation}
F_{\rm line}\!=\!\iint(I_{\nu}-I_{\rm cont})\,d\Omega\,d\nu \, \approx \, \!A\,B_{\nu}(T_{\rm gas})\!\int\!\phi(\nu)\,d\nu \,\,\, ,
\label{eqn:Fline}
\end{equation}
where $A$ is the disk surface area as seen by the observer.

If we assume a square line profile function in velocity space such that $\phi(\nu)\!=\!1$ if $-\frac{\Delta {\rm v}}{2} < {\rm v} < +\frac{\Delta {\rm v}}{2}$ and $\phi(\nu)\!=\!0$ otherwise, we can re-write equation~(\ref{eqn:Fline}) in the Rayleigh-Jeans limit as
\begin{equation}
F_{\rm line}\!=\!A\,B_{\nu}(T_{\rm gas})\,\Delta\nu\!=\!A\,2\,k\,T_{\rm gas}\,\Delta {\rm v}\frac{\nu}{c}^3
\end{equation}
using $\frac{\Delta {\rm v}}{c}\!=\!\frac{\Delta\nu}{\nu}$. For line ratios, the quantities $A$, $T_{\rm gas}$ and $\Delta {\rm v}$ are identical, and hence
\begin{equation}
\frac{F_{\rm line,1}}{F_{\rm line,2}}\!=\!\left( \frac{\nu_1}{\nu_2} \right)^3\,\,\, .
\end{equation}
The corresponding values for the three CO line ratios are overplotted with dashed lines on Fig.~\ref{fig:co_fluxes}, and agree indeed well with the limiting behaviour of the line ratios at high disk masses, $M_{\rm disk}\!\geq\!10^{-3}$. Thus, we conclude that CO line formation in high mass disks occurs under optically thick LTE conditions. At lower masses the line ratios deviate from these values since the assumption of optical thick LTE no longer holds.

Changing the effective temperature of the central star or changing the dust properties affects the CO lines to a lesser extent than the fine structure lines. Changes to the continuum and the lines are generally within 50\%. Even though the formation height of CO might change slightly, the excitation conditions are still close to LTE. While the CO mass in the models stays within a few percent of the original model (except in the ISM grain model, where it is 26\% smaller) the CO mass averaged temperature changes up to a factor two. The effect on the line fluxes correlates directly with the temperature change, with the higher temperatures giving systematically higher fluxes. At these radio wavelengths, the change in continuum emission amounts to less than 25\%, even between the models with different dust properties. The most extreme model is the one with an effective temperature of $10500$~K where the mass averaged CO temperature rises to 110~K and hence the line fluxes increase by $\sim 60-70$\%.

\begin{figure}
\centering
\hspace*{-1.cm}
\includegraphics[width=9cm]{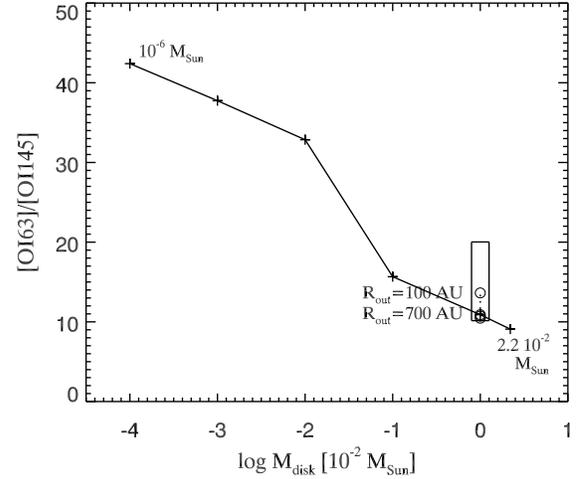} 
\vspace*{-3mm}
\caption{[O\,{\sc i}]\,63/145 $\mu$m line ratio as a function of disk gas mass. In the high disk mass regime, both lines are optically thick and the line ratio levels off around 10. At very low disk masses both lines are optically thin and the line ratio of $\sim40$ corresponds to gas temperatures above 100~K and a wide range of densities \citep[see][]{Tielens1985}. Plus signs denote the standard mass sequence, open circles models with varying outer radius (100, 300, 500 and 700~AU). The solid box show the range of line ratios that occurs with varying dust properties to the extremes, ISM type grains ($a_{\rm min}=0.05, a_{\rm max}=1~\mu$m) and large grains ($a_{\rm min}=1, a_{\rm max}=200~\mu$m).}
\label{fig:lineratiosOI}
\end{figure}

\begin{figure}
\centering
\hspace*{-1.cm}
\includegraphics[width=9cm]{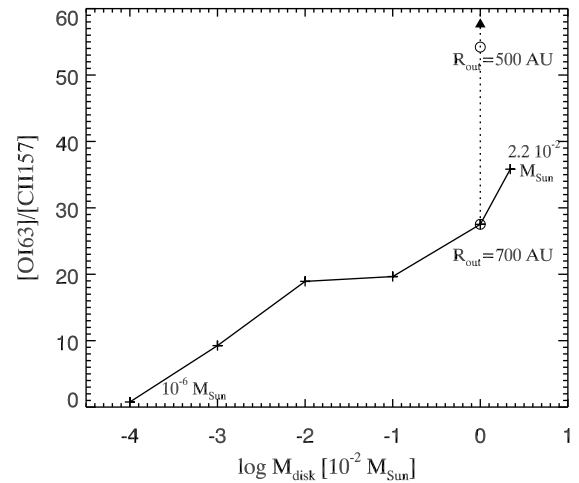} 
\vspace*{-3mm}
\caption{[O\,{\sc i}]\,63/[C\,{\sc ii}]\,158 $\mu$m line ratio as a function of disk gas mass. Plus signs denote the standard mass sequence, open circles models with varying outer radius (500 and 700~AU).}
\label{fig:lineratiosOICII}
\end{figure}

\begin{figure}
\centering
{
\hspace*{-1.cm}
\includegraphics[width=9cm]{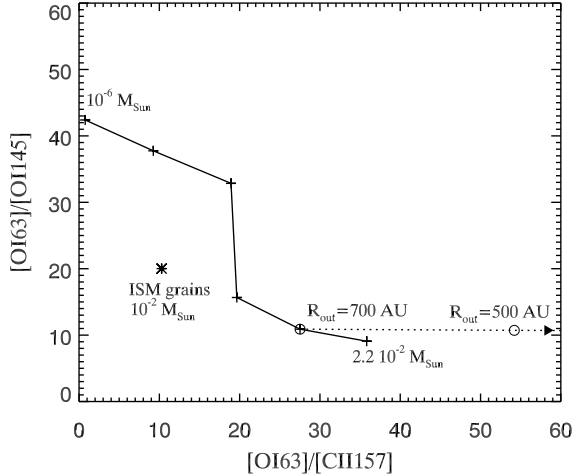} }
\vspace*{-3mm}
\caption{[O\,{\sc i}]\,63/145 $\mu$m versus [O\,{\sc i}]\,63/[C\,{\sc ii}]\,158 $\mu$m line ratio in the series of Herbig Ae disk models. Plus signs denote the standard mass sequence, open circles models with varying outer radius (500 and 700~AU). The data points for smaller outer radii fall well outside the plotting range as indicated by the arrow (see Table~\ref{tab:OIlineemission_models}). The star denotes the location of the $10^{-2}$~M$_\odot$ disk model with small grains ($a_{\rm min}=0.05, a_{\rm max}=1$~$\mu$m)}
\label{fig:lineratiosPDR}
\end{figure}

\section{Discussion}
\label{discussion}

The goal of this study is a first assessment of the diagnostic power of the fine structure lines of neutral oxygen, ionized carbon and of the CO submm lines in protoplanetary disks. This work prepares the ground for a systematic analysis of a much larger and more complete grid of protoplanetary disk models. Based on the detailed radiative transfer calculations of the previous section, we discuss here the potential of the various lines in the context of deriving physical disk properties such as gas mass, gas temperatures and disk extension.

\subsection{[O\,{\sc i}]\,63 line}

Using one dimensional modeling of photon-dominated regions (PDRs), \citet{Liseau2006} find that the total intensity of the $63$~$\mu$m line can be used as an indicator of the gas temperature in outflows from young stellar objects. From detailed two dimensional modeling of protoplanetary disks, \citet{Woitke2009a} conclude that the [O\,{\sc i}]\,63\,$\mu$m line alone will indeed provide a useful tool to deduce the gas temperature in the surface layers of protoplanetary disks, especially the hot inner disk surface. However, spatially unresolved observations that measure only integrated [O\,{\sc i}]\,63\,$\mu$m line fluxes in sources at typical distances of 140~pc will not be able to detect the presence of hot gas very close to the star (inside $\sim 10$~AU). The total line flux is dominated by emission coming from the cooler outer regions (Table~\ref{tab:OIlineemission_models}).

\subsection{[O\,{\sc i}]\,63/145 line ratio}

The diagnostic power of the [O\,{\sc I}]\,63/145\,$\mu$m line ratio has been discussed by \citet{Tielens1985} for PDRs and by \citet{Liseau2006} for outflows around young stellar objects (YSOs). Very low ratios of a few can be explained if both lines are optically thick \citep[see Fig.~2 of][]{Tielens1985}. In that case the diagnostic power of this line ratio is very limited.  \citet{Liseau2006} find that even in the optically thin limit, the line ratio can be degenerate with low temperature molecular gas giving the same result as high temperature atomic gas. This is due to the different collisional cross sections for O-H and O-H$_2$. From ISO observations, \citet{Lorenzetti2002} found ratios between 2 and 10 for Herbig Ae/Be stars. They derive from PDR modeling of a clumpy medium that the low line ratios of a few would require very high column densities of $N_\HH \sim 10^{22}$~cm$^{-3}$ ($A_V \sim 18 - 27$~mag) and conclude that better models of the emitting region are required for the physical interpretation of this data. Studying the outflows of several hundred YSOs with ISO, \citet{Liseau2006} found [O\,{\sc I}]\,63/145\,$\mu$m line ratios between $\sim 1$ and 50. They suggest that confusion with relatively cool but optically thick envelope gas ($T < 100$~K) could explain the low line ratio in some cases. 

Fig.~\ref{fig:lineratiosOI} shows that the [O\,{\sc i}]\,63/145 line ratio increases towards lower masses (10 for $\sim 10^{-2}$~M$_\odot$ to 40 for  M$_{\rm disk} < 10^{-4}$~M$_\odot$). Our line ratio for massive disks corresponds well to the expected value for both lines being optically thick and arising from relatively warm gas, $T > 100$~K, that resides in the surface layers of the disk. As the disk mass decreases, first the 145 and then the 63~$\mu$m line become optically thin. Eventually, the optically thin limit is reached for disk masses smaller than $10^{-4}$~M$_\odot$, leading to line ratios of $\sim 40$. In a simple one dimensional model, this corresponds to gas temperatures in excess of 100~K and a wide range of densities. The line profiles show an increasing peak separation towards lower mass disks (see Fig.~\ref{fig:OI63}). In fact, the line emission in low mass disks is dominated by disk material close to the star ($r<100$~AU), while in high mass disks, the emission comes from the disk surface out to $\sim 200$~AU and is hence dominated by the outer regions ($30 < r < 200$~AU). The sequence of the $10^{-2}$~M$_\odot$ disk model with varying outer radii (dotted line in Fig.~\ref{fig:lineratiosOI}) shows that the [O\,{\sc i}]\,63/145 line ratio is indeed not sensitive to the disk outer radius unless the disk starts to become smaller than the region where most emission originates (see the $R_{\rm out}=100$ AU model).

The effective temperature and thus amount of UV radiation from the central star has a negligible impact on the line ratio in the temperature range of Herbig Ae stars. Both lines increase by the same amount (up to a factor 7 for the individual line fluxes for $T_{\rm eff}=10\,500$~K compared to the standard case of $T_{\rm eff}=8460$~K) due to the overall warmer gas temperatures (the oxygen mass does not change by more than 5\%). The change in line ratio due to a factor 10 decrease in disk mass is much larger than the change due to a different effective temperature of the star.

On the other hand, the grain size distribution does have a larger effect on the line ratios as expected from the impact on the disk structure (see Sect.~\ref{physicalstructure:dustopacities}). The line ratio is 10\% smaller for the disk model with large micron sized dust grains. Using only small sub-micron sized grains (similar to the ISM composition) increases the line ratio by a factor two, thus mimicking a disk mass that is lower by more than a factor ten. Since the disk with small ISM grains is flatter and reaches thus higher particle densities, O depletes more readily into species such as water and water ice, hence decreasing the overall vertical column density of neutral oxygen (making the lines thus marginally optically thin). However, the grain size distributions chosen here present the extreme ends and are easily distinguished from looking at the SED. 

The typical hydrogen number densities in the emitting region range up to $10^8$~cm$^{-3}$, higher than those generally covered in PDR models (even in clumpy PDRs). The vertical column density of neutral oxygen reaches values of $10^{20}$~cm$^{-3}$ and temperatures above 200~K inside 10~AU and $10^{19}$~cm$^{-3}$ and temperatures below 100~K in the outer disk ($10^{-2}$~M$_\odot$ disk model). The large difference to PDR conditions can explain the problems and high extinction values that \citet{Lorenzetti2002} found from their classical PDR analysis. However, our highest mass disk models do not show line ratios as low as 2, the most extreme cases of the YSOs studied by \citet{Liseau2006}. For embedded objects, confusion with relatively cool but optically thick envelope gas ($T < 100$~K) is still a viable explanation of  the low ratios.

\subsection{[C\,{\sc ii}]\,158\,$\mu$m line}

The [C\,{\sc ii}]\,158~$\mu$m line emission is a very sensitive tracer of the outer disk radius. We come back to this point in Sect.~\ref{diag:COsubmm}. As expected, the integrated line flux from our models scales as $R_{\rm out}^2$. This makes it more difficult to detect smaller size disks that are only of the order of 100~AU in size.

Due to its low excitation temperature, this line is also strongly affected by confusion with any type of diffuse and molecular cloud material in the line of sight. The problem can be addressed with Herschel/PACS by using the surrounding pixels on the detector ($5 \times 5$ pixels with the disk being spatially unresolved) for a reliable off-source flux determination.

Changing the dust properties to very large grains results in a much lower [C\,{\sc ii}]\,158~$\mu$m line flux, the reason being that above the $\chi/n_\HH=0.01$ layer outside of 100~AU, the gas temperature drops below the excitation temperature for this line. The gas inside 30~AU becomes much warmer due to the gas-dust de-coupling, thereby generating a more pronounced secondary rim. This casts an efficient shadow on the outer disk, thereby enhancing the CO formation (H$_2$ and CO self-shielding) and thus the molecular line cooling.

\subsection{[O\,{\sc i}]\,63/[C\,{\sc ii}]\,158 line ratio}

The [O\,{\sc i}]\,63/[C\,{\sc ii}]\,158 ratio depends even more on disk mass than the [O\,{\sc i}]\,63/145 line ratio. It would thereby be a prime diagnostic for relative disk mass estimates. However, while the ratio of the two oxygen lines does not change very much with disk outer radius, the [O\,{\sc i}]\,63/[C\,{\sc ii}]\,158 ratio does depend strongly on it (Fig.~\ref{fig:lineratiosOICII}). The $10^{-2}$~M$_\odot$ disk model with an outer radius of 100~AU has a more than 10 times higher [O\,{\sc i}]\,63/[C\,{\sc ii}]\,158 line ratio than the same model with an outer radius of 700~AU. 

Because of this degeneracy with disk size, the diagnostic value of this ratio alone is limited for individual objects. Its use would require the a-priori knowledge of many other disk parameters such as the disk outer radius, grain sizes and composition. Rather we suggest that it could be used in a statistical way on a large sample of disks that share the same classification group, e.g. based on their IR SEDs. Alternatively, a combination of the [O\,{\sc i}]\,63/145 and [O\,{\sc i}]\,63/[C\,{\sc ii}]\,158 line ratios could be used to break the degeneracy with disk size. Fig.~\ref{fig:lineratiosPDR} shows clearly that disks with a smaller outer radius, e.g. 300~AU, lie well to the right of the standard relation, while the presence of an inner hole for example (30~AU size here) does not affect the location in this diagnostic diagram.

As for the oxygen fine structure line ratio, the [O\,{\sc i}]\,63/[C\,{\sc ii}]\,158 ratio depends strongly on the chosen grain size distribution. Additional SED observations are required to constrain the dust parameters to a reasonable range. It is important to note that the change in line ratios in the [O\,{\sc i}]\,63/145 versus [O\,{\sc i}]\,63/[C\,{\sc ii}]\,158 plot has a direction different from what we expect from a change in disk mass. Also, the [O\,{\sc i}]\,63/[C\,{\sc ii}]\,158 ratio is less sensitive to the effective temperature of the star than it is to disk mass.

\subsection{CO submm lines and line ratios}
\label{diag:COsubmm}

The low excitation CO rotational lines are often used to prove the existence of primordial gas in
protoplanetary disk and in combination with optically thin line of isotopologues to measure the total gas mass \citep[e.g.][]{Thi2001, Panic2008}. Due to the high Einstein $A$ coefficients and high abundances of $^{12}$CO in molecular regions, the low rotational CO lines become optically thick at low $A_V$. Thus, they originate mainly in the optically thin disk surface \citep{vanZadelhoff2001}, where gas and dust temperature decouple according to our models. However, as in the case of the [C\,{\sc ii}] line, CO submm lines can be contaminated by low density cold remnant cloud material. The amount of contamination can be estimated either from offset positions for single dish observations or through interferometry.

Isotopologues of CO have much lower abundances of $^{12}$CO/$^{13}$CO $=77$ in the ISM \citep{Wilson1994} and even higher values of 100 in disks \citep[selective photodissociation; see for most recent work][]{Smith2009, Woods2009, Visser2009}. Thus, they can probe deeper into the disk and are --- in the absence of freeze-out --- excellent tools for measuring the disk mass. \citet{Panic2008} use for example the optically thin $^{13}$CO J=2-1 line from the disk around the Herbig star HD\,169142 to infer a disk mass of $0.6 - 3.0 \,10^{-2}$~M$_\odot$. 

Our analysis shows indeed a relatively flat relationship for the $^{12}$CO lines with disk gas mass, confirming thus that the low rotational CO lines by themselves are not a good tracer of gas mass (Fig.~\ref{fig:co_fluxes}). The lines become optically thick even at very low disk masses and hence a reliable inversion of the line fluxes to infer gas mass proves impossible. Similarly there are degeneracies in the computed line ratios J=2-1/1-0, J=3-2/1-0 and J=3-2/2-1 which suggest limitations in the diagnostic power of these line ratios.

The CO lines can be used to measure independently the outer disk radius (Sect.~\ref{COlineform}) and thus mitigate the problem of employing the [C\,{\sc ii}]\,158 line for disk diagnostics. \citet{Dent2005} and \citet{Panic2009} demonstrate the power of single dish CO observations in deriving disk outer radii and inclinations, thus constraining SED modeling that is rather insensitive to $R_{\rm out}$ and degenerate for inclination and inner radius. \citet{Hughes2008a} note that the outer radius derived from dust continuum observations is always smaller than that derived from CO emission lines. Even though they can explain this within homogeneous gas and dust models ($R_{\rm out}({\rm gas}) = R_{\rm out}({\rm dust})$ and a realistic outer edge, i.e. an exponential density profile), this finding reveals the problems in mixing quantities deduced from dust and gas observations. Since C$^+$ and CO arise both from the gas in the outer disk, we do not need to rely on dust observations, but can instead use a gas tracer to measure the gas outer disk radius. 

The [C\,{\sc ii}] line probes in general slightly higher layers than the CO lines; hence the model results suggest that we can trace the vertical gas temperature gradient in disks. Table~\ref{tab:masses} shows that for the most massive disks, $\langle T_{\rm C^+} \rangle$ is 50\% larger than $\langle T_{\rm CO} \rangle$.

\subsection{Comparison with observations for MWC480}

Since we used the Herbig star MWC480 to motivate our choice of parameters, we briefly discuss the line emission from the models together with observations available from the literature. The goal is not to obtain a best-fitting model, but to provide an impression of the model results within the observational context. Table~\ref{tab:MWC480} gives an overview of selected observational data for MWC480 (detections and upper limits) from the literature. The CO mass derived from $^{13}$CO 3-2 is $1.7\,10^{-4}$~M$_\odot$ and the total gas mass derived from the 1.3~mm continuum flux is $2.2\pm 1.0 \,10^{-2}$~M$_\odot$ \citep{Thi2001}. The $^{12}$CO and $^{13}$CO lines suggest an outer disk radius of $\sim 750$ and $\sim 500$~AU, respectively \citep{Pietu2007}.

The oxygen fine structure lines are not detected with the ISO satellite \citep{Creech2002}. Assuming the ISO instrumental resolution as FWHM of a potential line (0.3~$\mu$m for $\lambda<90~\mu$m, 0.6~$\mu$m for $\lambda>90~\mu$m), the ISO $3\sigma$ upper limits for the [O\,{\sc i}] lines can be derived from
\begin{equation}
F_{\rm upper~limit} \approx 3 \sigma \sqrt{0.5 \pi} \,\, {\rm FWHM}~{\rm W/m^2}\,\,\,.
\end{equation}
This gives values of $4.14\,10^{-16}$ and $1.35\,10^{-16}$~W/m$^2$ for the 63 and 145\,$\mu$m line, respectively. The upper limit for the 63\,$\mu$m line is only a factor 1.6 smaller than the flux predicted from the $10^{-3}$~M$_\odot$ disk model. The 145\,$\mu$m upper limit is less sensitive and hence even consistent with our $10^{-2}$~M$_\odot$ disk model. Changing the outer disk radius to 500~AU does not affect the [O\,{\sc i}] lines.

The observed [C\,{\sc ii}] line is much stronger than in any of our models, suggesting indeed significant contamination from diffuse background material in the large ISO beam (80").

Our $2.2\,10^{-2}$~M$_\odot$ model overpredicts the CO J=3-2 line flux by approximately one order of magnitude. By reducing the maximum grain size, the total grain surface area increases and dust and gas temperatures couple in the region where this CO line forms; thus yielding $\langle T_{\rm g} \rangle \sim 40$~K and a factor two lower line fluxes. Another factor two can each be gained by reducing the total disk mass and the disk outer radius to 500 instead of 700~AU.

The observed continuum flux at 1.3~mm is a factor 4 smaller than computed from the $2.2\,10^{-2}$~M$_\odot$ disk model. Reducing again either the total disk mass or the maximum grain size, reduces the modeled continuum fluxes to the right value.

To conclude, by tuning the input parameters, the models can get close to the observed continuum as well as line fluxes, thus providing additional confidence that the models capture the essential physics and chemistry of protoplanetary disks. Hence, besides studying physics and chemistry of protoplanetary disks in a more general context, these models can also be used in the detailed analysis of individual objects that possess a larger set of continuum and line observations. 

\begin{table}
\caption{Observed fluxes for MWC480}
\begin{tabular}{lcll}
Type & $\lambda$ & Flux & Reference \\
\hline
O\,{\sc i} & 63~$\mu$m & $\leqslant 1.1\,10^{-15}$ W/m$^2$/$\mu$m & (1) \\[2mm]
O\,{\sc i} & 145~$\mu$m & $\leqslant 1.8\,10^{-16}$ W/m$^2$/$\mu$m & (1) \\[2mm]
C\,{\sc ii} & 158~$\mu$m & $5.9\pm 0.6\,10^{-16}$ W/m$^2$ & (1) \\
 & & $5.5\pm 0.4\,10^{-16}$ W/m$^2$ & (2) \\[2mm]
CO  &  345 GHz & 2.88 K km/s & (3) \\
    &   & $4.23\,10^{-19}$~W/m$^2$ & (*) \\[2mm]
cont. &  1.3 mm & $360\pm24$ mJy & (4) \\
cont. &  2.7 mm & $39.9\pm 0.8$ mJy & (4) \\
\hline\\[1mm]
\end{tabular}
\hspace*{0mm}\begin{minipage}{9cm}
\footnotesize
References: (1) \citet{Creech2002}, (2) \citet{Lorenzetti2002}, (3) \citet{Thi2004}, (4) \citet{Mannings1997a}.\\[2mm]
(*) $\int T_{\rm mb}\,dv = 10^{-2} \, \lambda_{\rm cm}/(2k) \, (\Omega_a)^{-1}  \int F_\nu({\rm W/m^2/Hz})\,d\nu$, with $\Omega_a$ being the solid angle of the beam, i.e. $\pi (13.7"/2)^2$ for the JCMT.
\end{minipage}
\label{tab:MWC480}
\end{table}

\section{Conclusions}

We used here a limited series of Herbig Ae disk models computed with the new thermo-chemical disk code called {\sc ProDiMo} to study the origin and diagnostic value of the gas line tracers [C\,{\sc ii}], [O\,{\sc i}] and CO. We do not include in this study effects of X-ray irradiation, grain settling or mixing. Even though X-rays are generally of minor importance for Herbig stars, grain settling and large scale mixing processes could affect the conclusions. The effect of grain settling will be included in a forthcoming larger model grid, but mixing processes need to be addressed with dynamical time-dependant models. The Monte-Carlo radiative transfer code {\sc Ratran} is used to compute line profiles and integrated emission from various gas lines. 

The main results are:

\begin{itemize}
\item The [C\,{\sc ii}] line originates in the disk surface layer where gas and dust temperatures are decoupled. The total line strength is dominated by emission from the disk outer radius. Thus the 157.7~$\mu$m line probes mainly the disk extension and outer disk gas temperature. The line forms in LTE.
\item The [O\,{\sc i}] lines originate also in the disk surface layer even though somewhat deeper than the [C\,{\sc ii}] line. The main contribution comes from radii between 30 and 100~AU. The [O\,{\sc i}] lines form partially under NLTE conditions. Differences in line emission from escape probability and Monte Carlo techniques are smaller than 10\%.
\item CO submm lines are optically thick down to very low disk masses of $<10^{-4}$~M$_\odot$ and form mostly in LTE. $T_{\rm gas}=T_{\rm dust}$ is not a valid approximation for these lines. Differences in line emission from escape probability and Monte Carlo techniques are smaller than 3\%, except in the case of very optically thin disk models ($10^{-5}$ and $10^{-6}$~M$_\odot$).
\item The [O\,{\sc i}]\,63/145 $\mu$m and [O\,{\sc i}]\,63/[C\,{\sc ii}]\,158 $\mu$m line ratios trace disk mass in the regime between $10^{-2}$ and $10^{-6}$~M$_\odot$. Since the [C\,{\sc ii}]\,158 $\mu$m line is very sensitive to the outer disk radius, the [O\,{\sc i}]\,63/[C\,{\sc ii}]\,158 $\mu$m is degenerate in that respect and its use requires additional constraints from ancilliary gas and/or dust observations. The sensitivity of these two line ratios to the dust grain sizes underlines the importance of using SED constraints along with the gas modeling to mitigate the uncertainty of dust properties.
\item A combination of the [O\,{\sc i}]\,63/145 $\mu$m and [O\,{\sc i}]\,63/[C\,{\sc ii}]\,158 $\mu$m line ratios can be used to diminish the degeneracy caused by an unknown outer disk radius.
\item Neither total CO submm line fluxes nor line ratios can be used to measure the disk mass. However, the low rotational lines studied here provide an excellent tool to measure the disk outer radius and can thus help to mitigate the degeneracy between gas mass and outer radius found for the [O\,{\sc i}]\,63/[C\,{\sc ii}]\,158 $\mu$m line ratio.
\end{itemize}

\begin{acknowledgements}
We thank Dieter Poelman for fruitful discussions about radiative transfer methods and detailed code comparisons.
\end{acknowledgements}

\bibliography{reference}

\begin{thebibliography}{59}
\expandafter\ifx\csname natexlab\endcsname\relax\def\natexlab#1{#1}\fi

\bibitem[{Auer(1984)}]{Auer1984}
Auer, L. 1984, in Numerical Radiative Transfer, ed. W.~Kalkhofen (Cambridge
  Univ. Press, Cambridge), 101

\bibitem[{{Beckwith} {et~al.}(1986){Beckwith}, {Sargent}, {Scoville}, {Masson},
  {Zuckerman}, \& {Phillips}}]{Beckwith1986}
{Beckwith}, S., {Sargent}, A.~I., {Scoville}, N.~Z., {et~al.} 1986, \apj, 309,
  755

\bibitem[{{Beckwith} \& {Sargent}(1991)}]{Beckwith1991}
{Beckwith}, S.~V.~W. \& {Sargent}, A.~I. 1991, \apj, 381, 250

\bibitem[{{Bell} {et~al.}(1998){Bell}, {Berrington}, \& {Thomas}}]{Bell1998}
{Bell}, K.~L., {Berrington}, K.~A., \& {Thomas}, M.~R.~J. 1998, \mnras, 293,
  L83

\bibitem[{Bratley {et~al.}(1994)Bratley, Fox, \& Niederreiter}]{Bratley1994}
Bratley, P., Fox, B.~L., \& Niederreiter, H. 1994, {ACM} Transactions on
  Mathematical Software, 20, 494

\bibitem[{{Brott} \& {Hauschildt}(2005)}]{Brott2005}
{Brott}, I. \& {Hauschildt}, P.~H. 2005, in ESA Special Publication, Vol. 576,
  The Three-Dimensional Universe with Gaia, ed. C.~{Turon}, K.~S. {O'Flaherty},
  \& M.~A.~C. {Perryman}, 565--+

\bibitem[{{Chu} \& {Dalgarno}(1975)}]{Chu1975}
{Chu}, S.-I. \& {Dalgarno}, A. 1975, Royal Society of London Proceedings Series
  A, 342, 191

\bibitem[{{Creech-Eakman} {et~al.}(2002){Creech-Eakman}, {Chiang}, {Joung},
  {Blake}, \& {van Dishoeck}}]{Creech2002}
{Creech-Eakman}, M.~J., {Chiang}, E.~I., {Joung}, R.~M.~K., {Blake}, G.~A., \&
  {van Dishoeck}, E.~F. 2002, \aap, 385, 546

\bibitem[{{Dartois} {et~al.}(2003){Dartois}, {Dutrey}, \&
  {Guilloteau}}]{Dartois2003}
{Dartois}, E., {Dutrey}, A., \& {Guilloteau}, S. 2003, \aap, 399, 773

\bibitem[{{Dent} {et~al.}(2005){Dent}, {Greaves}, \& {Coulson}}]{Dent2005}
{Dent}, W.~R.~F., {Greaves}, J.~S., \& {Coulson}, I.~M. 2005, \mnras, 359, 663

\bibitem[{{Draine} \& {Lee}(1984)}]{Draine1984}
{Draine}, B.~T. \& {Lee}, H.~M. 1984, \apj, 285, 89

\bibitem[{{Dutrey} {et~al.}(1997){Dutrey}, {Guilloteau}, \&
  {Guelin}}]{Dutrey1997}
{Dutrey}, A., {Guilloteau}, S., \& {Guelin}, M. 1997, \aap, 317, L55

\bibitem[{{Flower} \& {Launay}(1977)}]{Flower1977}
{Flower}, D.~R. \& {Launay}, J.~M. 1977, Journal of Physics B Atomic Molecular
  Physics, 10, 3673

\bibitem[{Hickernell \& Yue(2000)}]{Hickernell2000}
Hickernell, F.~J. \& Yue, R.-X. 2000, SIAM Journal on Numerical Analysis, 38,
  1089

\bibitem[{{Hogerheijde} \& {van der Tak}(2000)}]{Hogerheijde2000}
{Hogerheijde}, M.~R. \& {van der Tak}, F.~F.~S. 2000, \aap, 362, 697

\bibitem[{{Hughes} {et~al.}(2008{\natexlab{a}}){Hughes}, {Wilner}, {Kamp}, \&
  {Hogerheijde}}]{Hughes2008b}
{Hughes}, A.~M., {Wilner}, D.~J., {Kamp}, I., \& {Hogerheijde}, M.~R.
  2008{\natexlab{a}}, \apj, 681, 626

\bibitem[{{Hughes} {et~al.}(2008{\natexlab{b}}){Hughes}, {Wilner}, {Qi}, \&
  {Hogerheijde}}]{Hughes2008a}
{Hughes}, A.~M., {Wilner}, D.~J., {Qi}, C., \& {Hogerheijde}, M.~R.
  2008{\natexlab{b}}, \apj, 678, 1119

\bibitem[{{Isella} {et~al.}(2007){Isella}, {Testi}, {Natta}, {Neri}, {Wilner},
  \& {Qi}}]{Isella2007}
{Isella}, A., {Testi}, L., {Natta}, A., {et~al.} 2007, \aap, 469, 213

\bibitem[{{Jaquet} {et~al.}(1992){Jaquet}, {Staemmler}, {Smith}, \&
  {Flower}}]{Jaquet1992}
{Jaquet}, R., {Staemmler}, V., {Smith}, M.~D., \& {Flower}, D.~R. 1992, Journal
  of Physics B Atomic Molecular Physics, 25, 285

\bibitem[{{Jonkheid} {et~al.}(2007){Jonkheid}, {Dullemond}, {Hogerheijde}, \&
  {van Dishoeck}}]{Jonkheid2007}
{Jonkheid}, B., {Dullemond}, C.~P., {Hogerheijde}, M.~R., \& {van Dishoeck},
  E.~F. 2007, \aap, 463, 203

\bibitem[{{Juvela}(1997)}]{Juvela1997}
{Juvela}, M. 1997, \aap, 322, 943

\bibitem[{{Kamp} {et~al.}(2005){Kamp}, {Dullemond}, {Hogerheijde}, \&
  {Enriquez}}]{Kamp2005}
{Kamp}, I., {Dullemond}, C.~P., {Hogerheijde}, M., \& {Enriquez}, J.~E. 2005,
  in IAU Symposium, Vol. 231, Astrochemistry: Recent Successes and Current
  Challenges, ed. D.~C. {Lis}, G.~A. {Blake}, \& E.~{Herbst}, 377--386

\bibitem[{{Kamp} \& {van Zadelhoff}(2001)}]{Kamp2001}
{Kamp}, I. \& {van Zadelhoff}, G.-J. 2001, \aap, 373, 641

\bibitem[{{Koerner} {et~al.}(1993){Koerner}, {Sargent}, \&
  {Beckwith}}]{Koerner1993}
{Koerner}, D.~W., {Sargent}, A.~I., \& {Beckwith}, S.~V.~W. 1993, Icarus, 106,
  2

\bibitem[{{Launay} \& {Roueff}(1977)}]{Launay1977}
{Launay}, J.~M. \& {Roueff}, E. 1977, \aap, 56, 289

\bibitem[{{Liseau} {et~al.}(2006){Liseau}, {Justtanont}, \&
  {Tielens}}]{Liseau2006}
{Liseau}, R., {Justtanont}, K., \& {Tielens}, A.~G.~G.~M. 2006, \aap, 446, 561

\bibitem[{{Lorenzetti} {et~al.}(2002){Lorenzetti}, {Giannini}, {Nisini},
  {Benedettini}, {Elia}, {Campeggio}, \& {Strafella}}]{Lorenzetti2002}
{Lorenzetti}, D., {Giannini}, T., {Nisini}, B., {et~al.} 2002, \aap, 395, 637

\bibitem[{{Mannings} \& {Emerson}(1994)}]{Mannings1994}
{Mannings}, V. \& {Emerson}, J.~P. 1994, \mnras, 267, 361

\bibitem[{{Mannings} {et~al.}(1997){Mannings}, {Koerner}, \&
  {Sargent}}]{Mannings1997b}
{Mannings}, V., {Koerner}, D.~W., \& {Sargent}, A.~I. 1997, \nat, 388, 555

\bibitem[{{Mannings} \& {Sargent}(1997)}]{Mannings1997a}
{Mannings}, V. \& {Sargent}, A.~I. 1997, \apj, 490, 792

\bibitem[{{Meijerink} {et~al.}(2008){Meijerink}, {Glassgold}, \&
  {Najita}}]{Meijerink2008}
{Meijerink}, R., {Glassgold}, A.~E., \& {Najita}, J.~R. 2008, \apj, 676, 518

\bibitem[{Niederreiter(1992)}]{Niederreiter1992}
Niederreiter, H. 1992, Random number generation and quasi-Monte Carlo methods
  (Philadelphia, PA, USA: Society for Industrial and Applied Mathematics)

\bibitem[{{Nomura} {et~al.}(2007){Nomura}, {Aikawa}, {Tsujimoto}, {Nakagawa},
  \& {Millar}}]{Nomura2007}
{Nomura}, H., {Aikawa}, Y., {Tsujimoto}, M., {Nakagawa}, Y., \& {Millar}, T.~J.
  2007, \apj, 661, 334

\bibitem[{{Ossenkopf} \& {Henning}(1994)}]{Ossenkopf1994}
{Ossenkopf}, V. \& {Henning}, T. 1994, \aap, 291, 943

\bibitem[{Owen(2003)}]{Owen2003}
Owen, A.~B. 2003, ACM Trans. Model. Comput. Simul., 13, 363

\bibitem[{{Pani{\'c}} \& {Hogerheijde}(2009)}]{Panic2009}
{Pani{\'c}}, O. \& {Hogerheijde}, M.~R. 2009, \aap, 0, in preparation

\bibitem[{{Pani{\'c}} {et~al.}(2008){Pani{\'c}}, {Hogerheijde}, {Wilner}, \&
  {Qi}}]{Panic2008}
{Pani{\'c}}, O., {Hogerheijde}, M.~R., {Wilner}, D., \& {Qi}, C. 2008, \aap,
  491, 219

\bibitem[{{Pi{\'e}tu} {et~al.}(2007){Pi{\'e}tu}, {Dutrey}, \&
  {Guilloteau}}]{Pietu2007}
{Pi{\'e}tu}, V., {Dutrey}, A., \& {Guilloteau}, S. 2007, \aap, 467, 163

\bibitem[{{Pi{\'e}tu} {et~al.}(2005){Pi{\'e}tu}, {Guilloteau}, \&
  {Dutrey}}]{Pietu2005}
{Pi{\'e}tu}, V., {Guilloteau}, S., \& {Dutrey}, A. 2005, \aap, 443, 945

\bibitem[{{Press} {et~al.}(2002){Press}, {Vetterling}, \&
  {Flannery}}]{Press2002}
{Press}, W.~H.and~{Teukolsky}, S.~A., {Vetterling}, W.~T., \& {Flannery}, B.~P.
  2002, {Numerical recipes in C++ : the art of scientific computing} (Cambridge
  University Press)

\bibitem[{{Rodmann} {et~al.}(2006){Rodmann}, {Henning}, {Chandler}, {Mundy}, \&
  {Wilner}}]{Rodmann2006}
{Rodmann}, J., {Henning}, T., {Chandler}, C.~J., {Mundy}, L.~G., \& {Wilner},
  D.~J. 2006, \aap, 446, 211

\bibitem[{{Schinke} {et~al.}(1985){Schinke}, {Engel}, {Buck}, {Meyer}, \&
  {Diercksen}}]{Schinke1985}
{Schinke}, R., {Engel}, V., {Buck}, U., {Meyer}, H., \& {Diercksen}, G.~H.~F.
  1985, \apj, 299, 939

\bibitem[{{Sch{\"o}ier} {et~al.}(2005){Sch{\"o}ier}, {van der Tak}, {van
  Dishoeck}, \& {Black}}]{Lambda2005}
{Sch{\"o}ier}, F.~L., {van der Tak}, F.~F.~S., {van Dishoeck}, E.~F., \&
  {Black}, J.~H. 2005, \aap, 432, 369

\bibitem[{Sloan(1993)}]{Sloan1993}
Sloan, I.~H. 1993, SIAM Review, 35, 680

\bibitem[{{Smith} {et~al.}(2009){Smith}, {Pontoppidan}, {Young}, {Morris}, \&
  {van Dishoeck}}]{Smith2009}
{Smith}, R.~L., {Pontoppidan}, K.~M., {Young}, E.~D., {Morris}, M.~R., \& {van
  Dishoeck}, E.~F. 2009, \apj, 701, 163

\bibitem[{Sobol \& Shukhman(2007)}]{Sobol2007}
Sobol, I.~M. \& Shukhman, B.~V. 2007, Math. Comput. Simul., 75, 80

\bibitem[{{Stelzer} {et~al.}(2006){Stelzer}, {Micela}, {Hamaguchi}, \&
  {Schmitt}}]{Stelzer2006}
{Stelzer}, B., {Micela}, G., {Hamaguchi}, K., \& {Schmitt}, J.~H.~M.~M. 2006,
  \aap, 457, 223

\bibitem[{{Thi} {et~al.}(2001){Thi}, {van Dishoeck}, {Blake}, {van Zadelhoff},
  {Horn}, {Becklin}, {Mannings}, {Sargent}, {van den Ancker}, {Natta}, \&
  {Kessler}}]{Thi2001}
{Thi}, W.~F., {van Dishoeck}, E.~F., {Blake}, G.~A., {et~al.} 2001, \apj, 561,
  1074

\bibitem[{{Thi} {et~al.}(2004){Thi}, {van Zadelhoff}, \& {van
  Dishoeck}}]{Thi2004}
{Thi}, W.-F., {van Zadelhoff}, G.-J., \& {van Dishoeck}, E.~F. 2004, \aap, 425,
  955

\bibitem[{{Tielens} \& {Hollenbach}(1985)}]{Tielens1985}
{Tielens}, A.~G.~G.~M. \& {Hollenbach}, D. 1985, \apj, 291, 747

\bibitem[{{van Dishoeck} {et~al.}(2008){van Dishoeck}, {Jonkheid}, \& {van
  Hemert}}]{vanDishoeck2008}
{van Dishoeck}, E.~F., {Jonkheid}, B., \& {van Hemert}, M.~C. 2008, ArXiv
  e-prints

\bibitem[{{van Zadelhoff} {et~al.}(2001){van Zadelhoff}, {van Dishoeck}, {Thi},
  \& {Blake}}]{vanZadelhoff2001}
{van Zadelhoff}, G.-J., {van Dishoeck}, E.~F., {Thi}, W.-F., \& {Blake}, G.~A.
  2001, \aap, 377, 566

\bibitem[{{Visser} {et~al.}(2009){Visser}, {van Dishoeck}, \&
  {Black}}]{Visser2009}
{Visser}, R., {van Dishoeck}, E.~F., \& {Black}, J.~H. 2009, ArXiv e-prints

\bibitem[{{Wilson} \& {Bell}(2002)}]{Wilson2002}
{Wilson}, N.~J. \& {Bell}, K.~L. 2002, \mnras, 337, 1027

\bibitem[{{Wilson} \& {Rood}(1994)}]{Wilson1994}
{Wilson}, T.~L. \& {Rood}, R. 1994, \araa, 32, 191

\bibitem[{{Woitke} {et~al.}(2009{\natexlab{a}}){Woitke}, {Kamp}, \&
  {Thi}}]{Woitke2009a}
{Woitke}, P., {Kamp}, I., \& {Thi}, W.-F. 2009{\natexlab{a}}, \aap, 501, 383

\bibitem[{{Woitke} {et~al.}(2009{\natexlab{b}}){Woitke}, {Thi}, {Kamp}, \&
  {Hogerheijde}}]{Woitke2009b}
{Woitke}, P., {Thi}, W.-F., {Kamp}, I., \& {Hogerheijde}, M.~R.
  2009{\natexlab{b}}, \aap, 501, L5

\bibitem[{{Woods} \& {Willacy}(2009)}]{Woods2009}
{Woods}, P.~M. \& {Willacy}, K. 2009, \apj, 693, 1360

\bibitem[{{Zuckerman} {et~al.}(1995){Zuckerman}, {Forveille}, \&
  {Kastner}}]{Zuckerman1995}
{Zuckerman}, B., {Forveille}, T., \& {Kastner}, J.~H. 1995, \nat, 373, 494

\end{thebibliography}


\end{document}